\documentclass[preprint,12pt]{elsarticle}

\usepackage{graphicx}%
\usepackage{multirow}%
\usepackage{amsmath,amssymb,amsfonts}%
\usepackage{amsthm}%
\usepackage{mathrsfs}%
\usepackage[title]{appendix}%
\usepackage{xcolor}%
\usepackage{textcomp}%
\usepackage{manyfoot}%
\usepackage{booktabs}%
\usepackage{algorithm}%
\usepackage{algorithmicx}%
\usepackage{algpseudocode}%
\usepackage{xcolor,graphicx,float}
\usepackage{epstopdf}
\usepackage{stfloats}
\usepackage[caption=false,font=normalsize,labelfont=sf,textfont=sf]{subfig}
\usepackage{listings}%
\usepackage{array}
\usepackage{url}
\usepackage{verbatim}
\usepackage{dutchcal}
\usepackage{widetext}
\usepackage{cases}
\usepackage{caption}
\usepackage{hyperref}
\allowdisplaybreaks[1]
\journal{Applied Mathematics Letters}

\begin{document}

\begin{frontmatter}



\title{Renormalization group based implicit function approach to connecting orbits}


\author{Pengfei Guo$^{1}$,  Yueheng Lan$^{1,2,3*}$, Jianyong Qiao$^{1,2*}$}
\address{1.School of science, Beijing University of Posts and Telecommunications, Beijing 100876, China;\\
2.Key Laboratory of Mathematics and Information Networks(Beijing University of Posts and Telecommunications), Ministry of Education, Beijing 100876, China;\\
3.State Key Lab of Information Photonics and Optical Communications, Beijing University of Posts and Telecommunications, Beijing 100876, China.}
\cortext[cor1]{Corresponding author. Email address: lanyh@bupt.edu.cn,qjy@bupt.edu.cn}
\begin{abstract}
Connecting orbits are important invariant structures in the state space of nonlinear systems and various techniques are designed for their computation. However, a uniform analytic approximation of the whole orbit seems rare. Here, based on renormalization group, an implicit function scheme is designed to effectively represent connections of disparate types, where coefficients of the defining function satisfy a set of  linear algebraic equations, which greatly simplifies their computation. Unknown system parameters are conveniently determined by minimizing an error function. Symmetry may be profitably utilized to reduce the computation load. Homoclinic or heteroclinic connections are found in five popular examples approximately or exactly, demonstrating the effectiveness of the new scheme.
\end{abstract}




\end{frontmatter}



\section{Introduction}\label{sec1}\par
Homoclinic or heteroclinic connections are vital geometric objects for investigating complex  dynamics, enabling comprehension of transitions from one state to another or delimiting boundary of different dynamical regimes.  Therefore, they play important roles in diverse fields including mathematics,  physics, chemistry, biomathematics, economics and so on. For example, solitons in water waves, optical transmission or reaction-diffusion systems, instantons in quantum mechanics could all be regarded as connecting orbits~\cite{1_1983Solitons,2_2000Nonlinear,3_1992Fronts,4_zwanzig2001nonequilibrium}.  Heteroclinic orbits are seen in the study of ecosystems to describe dynamic transitions of a species' population~\cite{5_2000Nonlinear} and of electronic circuits to assess bifurcations~\cite{8_khibnik1993periodic}. In meteorology dynamics, connections serve as transition paths of weather patterns between different states~\cite{9_1963Deterministic,10_Robinson2000Nonsymmetric,11_2013Hopf,12_Leonov2016Necessary}.  Hence, searching and computing these orbits are essential in the investigation of nonlinear dynamics.\par
Over a long period, numerical methods are most widely adopted means for computing the connecting orbits, including the  shooting technique~\cite{13_champneys1993hunting,44_hassard1994existence},  the variational approach~\cite{14_2014A},  adjoint-based methods~\cite{15_2019Computing}, and so on. In comparison, analytical approximations do not seem so popular although they can be very useful in checking the parameter dependence, fixing the existence condition and determining the asymptotic dynamics. Nevertheless, various approaches have been suggested recently to analytically approximate connections, such as techniques based on the center manifold \cite{19_1981Applications}, reductive perturbation methods~\cite{20_1984Chemical}, variational iterations~\cite{21_1999Variational}.  A.F.Vakakis and M.F.A.Azeez~\cite{16_1998Analytic} approximated the homoclinic orbit of the Lorenz system and improved the accuracy through Pad\'{e} approximation, where the solution was written as two infinite  series jointed at the initial point. Similar job was done by  J. Song~\cite{18_2019The}.  Symmetry is considered in the computation of heteroclinic orbits of L\"{u} system and Zhou's system~\cite{17_2012Existence} which are represented with piecewise continuous functions since the coefficients of the analytic expressions have to be determined by solving nonlinear algebraic equations, which may not be easy, especially in high dimensions since  the connecting orbits only exist at particular values of the parameters~\cite{16_1998Analytic,17_2012Existence,18_2019The}.\par
Recently, the renormalization group (RG) method~\cite{35_Yueheng2013Bridging} was used in the detection of connecting orbits.
The RG theory~\cite{24_Wilson1971Renormalization,25_1971Renormalization,26_Wilson1972Renormalization,27_1989Quantum} was first introduced in particle physics and later extended to statistical physics. It was found later~\cite{22_1994Renormalization,23_1995The} that the RG method enables an asymptotic analysis of both ordinary and partial differential equations,  which offers a reduction in certain types of dynamical systems~\cite{28_bricmont1992renormalization,29_bricmont1994renormalization}. Later, a multitude of other studies were conducted to validate the efficacy of the RG approach~\cite{34_kunihiro2022geometrical,33_Chiba2008Approximation} or to associate it to conventional analyzing techniques~\cite{30_2000On,31_deville2008analysis,32_chiba2008c}. The RG analysis~\cite{22_1994Renormalization,23_1995The} is quite straightforward and flexible and so fits for  computing of invariant objects, such as cycles, connections or other sets of higher dimensions. Although Lan approximated the connections quite well~\cite{35_Yueheng2013Bridging}, there are still problems that need to be solved. In the computation, asymptotic analysis is carried out only at one end. Thus the other end is only an approximation, which may not fit the true orbit well, especially when the orbit is not simple, such as the homoclinic or spiraling orbits. In recent years, the RG method has also been extended to the study of stochastic differential equations~\cite{36_kupiainen2016renormalization,37_2021Renormalization,38_guo2024renormalization},   time-delay differential equations~\cite{39_goto2007renormalization,40_2023Renormalization}, and so on.
\par
 In this paper, an implicit function approach is employed to overcome the difficulties mentioned above. In this formulation, the expansions at the two ends are considered on an equal footing to determine the coefficients in the relation function. Only a set of linear equations need to be solved for these coefficients as a function of system parameters which may enter the equation in a nonlinear way. If some of these parameters are unknown and have to be determined together with the connection, an error function is constructed, the minimization of which gives the required parameters values. The form of the relation function is designed according to the RG-based expansion at the ends of the orbit. In the presence of symmetry, the computation could be much simplified.  Five typical systems are used to demonstrate the new scheme, in which heteroclinic or heteroclinic orbits are found, even with quite complex structures like spirals.  In one case, exact solutions are computed with this implicit function approach.
\par
 The paper is organized as follows. Section \ref{sec2} explains how the RG method is applied to the solution of  differential dynamical systems, in particular, to the search for connecting orbits with the help of implicit functions. In section \ref{sec3}, five examples are used to illustrate the application of the proposed scheme.  Finally, conclusions are drawn and future directions are pointed out in Section \ref{sec4}.

\section{\textbf{General Algorithm}}\label{sec2}
\subsection{\textbf{The RG  scheme for differential equations}}\label{subsec2.1}
In this section, we explain the general RG scheme for solving nonlinear differential equations. Let's consider the following  $n$-d differential equation
\begin{equation}
	\begin{array}{lll}
\dot{X}=G(x)=LX+\epsilon N(X),
	\end{array} \label{2.1}
\end{equation}
where $X=(x_1,x_2,...,x_n)^\top$ denotes the state variable and $L=\mathrm{diag}(l_1,l_2,...,l_n)$ is an $n\times n$ diagonal matrix. Here for simplicity, without much loss of generality, we assume that $L$ is already diagonalized.  $N(x)$ represents a nonlinear vector function. $\epsilon$ is a small parameter which may be set to $1$ towards the end of the calculation. To solve $(\ref{2.1})$, we start from a regular expansion
\begin{equation}
	\begin{array}{lll}
X=X_1+\epsilon X_2+\epsilon^2 X_3+...
	\end{array} \label{2.2}
\end{equation}
the substitution of which into $(\ref{2.1})$ yields at different orders of $\epsilon$
\begin{equation}
	\begin{array}{lll}
\dot{X_1}&=&LX_1 \\
\dot{X_2}&=&LX_2+N(X_1) \\
&...,&
	\end{array} \label{2.3}
\end{equation}
which may solved sequentially.  For example, $X_1$ could be written as
\begin{equation}
	\begin{array}{lll}
X_1(t,t_0)=Ae^{L(t-t_0)},
	\end{array} \label{2.4}
\end{equation}
where $t_0$ is the initial time and  $A=A(t_0)=[A_1(t_0),A_2(t_0),...,A_n(t_0)]^{\top}$ is a constant vector representing the initial position. Substituting $(\ref{2.4})$ into the second equation of  $(\ref{2.3})$, we get
\begin{equation}
	\begin{array}{lll}
X_2(t,t_0)=\tilde{A}e^{L(t-t_0)}+e^{L(t-t_0)}\int_{t_0}^te^{-L(\tau-t_0)}N[Ae^{L(\tau-t_0)}]d\tau,
	\end{array} \label{2.5}
\end{equation}
where $\tilde{A}$ is an undetermined constant vector, which may be used to choose different parameterization for the approximate solution. In the current case,  as $L$ is diagonal,   each component of $X_2(t,t_0)$ could be computed individually. Putting the solutions at  different orders of $\epsilon$ back into the expansion $(\ref{2.2})$, we have the approximation $X=\tilde{X}[t;t_0,A(t_0)]$ which is supposedly valid only at small $\epsilon$.
\par

In case of autonomous equations, $t\,,t_0$ always appear as  $(t-t_0)$.  If there is a resonance, the term  $(t-t_0)$ becomes a prefactor before and is unbounded in the large $t$ limit, which renders invalid the expansion  $X=\tilde{X}[t;,t_0,A(t_0)]$. To subdue the difficulty, Chen and Goldenfeld.N.et.al \cite{22_1994Renormalization,23_1995The} introduced an intermediate time $\tau$ between $t$ and $t_0$ and used the integral constants $A$ to absorb terms containing $\tau-t_0$.  The independency of the solution on $\tau$ gives the renormalization equation for $A$.  In \cite{35_Yueheng2013Bridging,32_chiba2008c}, the process is much simplified with the following RG equation
\begin{equation}
\frac{d\tilde{X}[t;t_0,A(t_0)]}{dt_0}|_{t=t_0}=0 \,
 \label{2.6}
\end{equation}
from which $dA/dt_0$ is obtained, governing the evolution of $A$.
The condition $t = t_0$ makes the evolution independent of $t$.  However, if $\tilde{X}[t;t_0,A(t_0)]$ is exact, the evolution of $A$ depends only on $t_0$ and $t = t_0$ is unnecessary.
\par

In this manuscript, we  focus on the computation of  connecting orbits in differential dynamical systems. For that, one renormalization variable may be chosen to describe the orbit~\cite{35_Yueheng2013Bridging}. After the equation for $A(t_0)$ is solved,  the approximate analytical solution is represented by a set of  polynomials of $A(t_0)$, one for each component. However, if the RG equation is derived around one fixed point, the orbit will not necessarily reach the fixed point exactly at the other end~\cite{35_Yueheng2013Bridging} since the solution is only an approximation. To overcome this difficulty, in this paper, we carry out the expansion on both ends to ensure the accuracy near the equilibria. To accommodate the two expansions, relation functions of state variables are designed to depict the connecting orbits. On the other hand, simple polynomials are hard to represent complex orbits such as spiral connections.  We extend the representation to polynomials with non-integer powers. As a result, the whole solution curve is described with same analytic expressions in contrast to most other works~\cite{16_1998Analytic,18_2019The,17_2012Existence}. In addition, specific forms of expressions could be chosen to reflect generic symmetry of the solution.
\subsection{\textbf{Implicit representation of the connecting orbits}}\label{subsec2.2}\par
In the following, detailed explanation is given to compute the implicit representation for a connection by using implicit function scheme. With the expansion~(\ref{2.3}) near the origin (supposed to be one end of the orbit), the series solution is
\begin{equation}
	\begin{array}{lll}
x_1&=A_1e^{l_1(t-t_0)}+\epsilon(P_{11}A_1^2e^{2l_1(t-t_0)}+P_{12}A_2^2e^{2l_2(t-t_0)}\\
&+...+P_{13}A_1A_2e^{(l_1+l_2)(t-t_0)}+...)+...,\\
x_2&=A_2e^{l_2(t-t_0)}+\epsilon(P_{21}A_1^2e^{2l_1(t-t_0)}+P_{22}A_2^2e^{2l_2(t-t_0)}\\
&+...+P_{23}A_1A_2e^{(l_1+l_2)(t-t_0)}+...)+...,\\
&...,\\
x_n&=A_ne^{l_n(t-t_0)}+\epsilon(P_{n1}A_1^2e^{(2l_1(t-t_0)}+P_{n2}A_2^2e^{2l_2(t-t_0)}\\
&+...+P_{n3}A_1A_2e^{(l_1+l_2)(t-t_0)}+...)+...,\\
	\end{array} \label{new1}
\end{equation}
where $\{P_{ij}\}_{i,j=1,2,...}$ are the known coefficients derived by solving the linear differential equations in Eq. $(\ref{2.3})$.  For convenience, we may assume the real parts of the first $p$ eigenvalues are positive while those of the rest $n-p$ ones are negative. Thus, the unstable directions require $A_{p+1}=A_{P+2}=,...,=A_{n}=0$ and  the stable ones demand ${A}_{1}=A_{2}=...={A}_{p}=0$.  With the expression~(\ref{new1}), the RG equation $(\ref{2.6})$ gives $\frac{dA_i}{dt_0}=l_iA_i$, for $i=1,...,n$ and theirs solution are $A_i=C_ie^{l_it}$, where $C_i$'s are arbitrary. A homoclinic orbit should start from the unstable direction and comes back along the stable direction.  When dealing with a heteroclinic orbit, an expansion at another equilibrium $X_0=(x_{10}\,,x_{20}\,,\cdots\,,x_{n0})^T$ is needed.
\begin{equation}
	\begin{array}{lll}
x_1&=x_{10}+\hat{A}_1e^{\tilde{l}_1(t-t_0)}+\epsilon(\hat{P}_{11}\hat{A}_1^2e^{2\tilde{l}_1(t-t_0)}+\hat{P}_{12}\hat{A}_2^2e^{2\tilde{l}_2(t-t_0)}\\
&+...+\hat{P}_{13}\hat{A}_1\hat{A}_2e^{(\tilde{l}_1+\tilde{l}_2)(t-t_0)}+...)+...,\\
x_2&=x_{20}+\hat{A}_2e^{\tilde{l}_2(t-t_0)}+\epsilon(\hat{P}_{21}\hat{A}_1^2e^{2\tilde{l}_1(t-t_0)}+\hat{P}_{22}\hat{A}_2^2e^{2\tilde{l}_2(t-t_0)}\\
&+...+\hat{P}_{23}\hat{A}_1\hat{A}_2e^{(\tilde{l}_1+\tilde{l}_2)(t-t_0)}+...)+...,\\
&...,\\
x_n&=x_{n0}+\hat{A}_ne^{\tilde{l}_n(t-t_0)}+\epsilon(\hat{P}_{n1}\hat{A}_1^2e^{2\tilde{l}_1(t-t_0)}+\hat{P}_{n2}\hat{A}_2^2e^{2\tilde{l}_2(t-t_0)}\\
&+...+\hat{P}_{n3}\hat{A}_1\hat{A}_2e^{(\tilde{l}_1+\tilde{l}_2)(t-t_0)}+...)+...\\
	\end{array} \label{new2}
\,,
\end{equation}\\
where the constants $\hat{A}_i$ should be chosen according to the sign of the real part of the eigenvalue $\tilde{l}_i$ as above.
\par
For a single connection in the $n$-d space, we need $n-1$ relation functions. The simplest situation is depicted by a polynomial expression as below
\begin{equation}
	\begin{array}{lll}
F_{1}&=\epsilon(a_{1,1}x_1+a_{1,2}x_2+...+a_{1,n}x_n)\\
&+\epsilon^2(b_{1,1}x_{1}^2+b_{1,2}x_2^2+b_{1,3}x_1x_2+...)+...\\
&=0,\\
F_{2}&=\epsilon(a_{2,1}x_1+a_{2,2}x_2+...+a_{2,n}x_n)\\
&+\epsilon^2(b_{2,1}x_{1}^2+b_{2,2}x_2^2+b_{2,3}x_1x_2+...)+...\\
&=0,\\
&...,\\
F_{n-1}&=\epsilon(a_{n-1,1}x_1+a_{n-1,2}x_2+...+a_{n-1,n}x_n)\\
&+\epsilon^2(b_{n-1,1}x_{1}^2+b_{n-1,2}x_2^2+b_{n-1,3}x_1x_2+...)+...\\
&=0,
	\end{array} \label{new3}
\end{equation}
where $\{F_i\}_{i=1,2,...,n-1}$ are relation functions the zeros of which that define a set of implicit functions.  $\{a_{i,j}\}_{i,j=1,2,...}$ are coefficients to be determined by matching the expansions on both ends of the connection. More explicitly, Substituting Eqs. $(\ref{new1})$ or $(\ref{new2})$ into relation functions $(\ref{new3})$ and comparing the coefficients of different orders of $A$ or $\hat{A}$, we obtain a set of linear equations for the coefficients $\{a_{i,j}\}_{i,j=1,2,...}$.  In $2d$ system, one relation function is sufficient while in $3d$ system, the connecting orbit is the intersection of the two surfaces described by two relation functions. To avoid degeneracy, the pick-up of proper  functional forms of relation functions is  crucial, which will be demonstrated in the examples below.
\par
\subsection{Generalized polynomials in multi-dimensions}
In high-dimensional systems, an equilibrium typically possesses multiple stable and unstable manifolds. In this case, sometimes it is necessary to determine the locus on these manifolds along which the connecting orbit leaves or reaches the equilibrium. In terms of the relation functions, it is easy to know the inter-dependence of the coordinate variables through local analysis but with possible unknown parameters that describe multiple stable or unstable directions and  can only be determined globally.  For example, if there are two unstable directions, we have $A_1=C_1e^{l_1t}\,, A_2=C_2e^{l_2t}$ with $\mathrm{Re} (l_1)>0\,,\mathrm{Re}(l_2)>0$. It is not hard to write down  $A_1=RA_{2}^{\alpha}$, with $\alpha=\frac{l_1}{l_2}$ and  $R$ is an unknown parameter used to select the filament that embeds the connecting orbit on the unstable manifolds. If we set $t=t_0$ in Eq.~(\ref{new1}), the original coordinates $x_1\,,x_2\,,\cdots, x_n$ are expressed as a function of $A_1, A_2$. Locally, $A_1\,,A_2$ could be written as a function of $x_1\,,x_2$ and therefore an relation function of $x_1\,,x_2$ is derived since $A_1=RA_{2}^{\alpha}$.  As $\alpha$ is not an integer in general, the relation function is a generalized polynomial with non-integer powers. One such example is given by Eq.~(\ref{3.7}) in the first example in Section~\ref{sec3}. To express spiral connections as in the second or the third example, the power $\alpha$ could even be a complex number!
\par
The parameter $R$ enters the relation function in a nonlinear way, which complicates the solution process.   Additionally,  the same difficulty also arises if there are other parameters in the equation that need to be determined.  To cope with this problem, we propose the following scheme  to estimate those parameters by minimizing the following error function
\begin{equation}
	\begin{array}{lll}
\Delta(R)=\Sigma_k^m\Sigma_{i}^{n-1}(\Sigma_j^{n}\frac{\partial F_i(k)}{\partial x_j}\dot{x}_j(k))^2,
\end{array} \label{new5}
\end{equation}
where $\{F_{i}\}_{i=1,2,...,n-1}$ are defined in Eq.~(\ref{new3}) which depend on the coefficients $\{a_{i,j}\,,b_{i,j}\,,\cdots\}$ in a linear way but the unknown parameters, say $R$, enters the equation  nonlinearly. In Eq.~(\ref{new5}), $\{\dot{x}_{j}\}_{j=1,2,...,n}$ is the original vector field $(\ref{2.1})$.   The equality $\Delta(R)=0$ indicates that the intersection of the surfaces defined by $\{F_{i}=0\}$ is a solution of the original equation which with proper boundary conditions delivers the connecting orbit.  Therefore, a minimization of  $\Delta$ drives the intersection to the right trajectory. In practice, the coefficients could be easily expressed with $R$ by solving a linear equation. That is why we write $\Delta=\Delta(R)$. Disparate minimization procedures could then be invoked to do this job started with an initial guess $R_0$. To reduce the computation load, the number of  the sampling points ${x_j(k)}$ could be small. In the presence of multiple unknown parameters, say $(R,S,T)$, similar procedure could be used to minimize the error function $\Delta=\Delta(R,S,T)$ as long as this function can be conveniently evaluated~\cite{press2007numerical}.

\subsection{Visualization}
To plot the intersection of multiple surfaces in high-dimensional phase space, an evolution equation is designed with the help of the implicit functions $(\ref{new3})$
\begin{equation}
	\begin{array}{lll}
&\sum^{n}_{i=1}\frac{\partial F_{1}}{\partial x_i}\dot{\tilde{x}}_i=0,\\
&\sum^{n}_{i=1}\frac{\partial F_{2}}{\partial x_i}\dot{\tilde{x}}_i=0,\\
&...,\\
&\sum^{n}_{i=1}\frac{\partial F_{n-1}}{\partial x_i}\dot{\tilde{x}}_i=0,\\
&\sum^{n}_{i=1}G_{i} \dot{\tilde{x}}_i=\sum^{n}_{i=1}G_{i}^2,
	\end{array} \label{new4}
\end{equation}
where $\{G_{i}\}_{i=1,2,...,n}$ are the velocity components of the original system $(\ref{2.1})$ and the variables $\{\tilde{x}_i\}_{i=1,2,...,n}$ satisfy Eqs. $(\ref{new3})$. Obviously, the first $n-1$ equations in system $(\ref{new4})$ define the orbit trajectory and the last equation tries to keep the magnitude of the original speed. Numerically, Eq.~(\ref{new4}) may be easily inverted to give  $\{\dot{\tilde{x}}\}_{i=1,2,...,n}$, which is then used to move on the connecting orbit.
\section{\textbf{Applications}}\label{sec3}
In this section, we explicitly show how the implicit function scheme explained above works in different dynamical systems. Analytic approximation in five typical examples with homoclinic or heteroclinic orbits will be provided.  The first example is a $2d$ differential dynamical system from biophysics, in which one heteroclinic orbit is very well represented with an implicit function involving non-integer powers. The second example is a  $2d$  system with one tunable parameter, the change of which leads to  different connections.   In one case,  a heteroclinic orbit  with a spiral structure appear which is  represented by an implicit function with complex powers, a little more involved than the first example.  A homoclinic and a heteroclinic orbit in the Lorenz system will be discussed in the third example where invariant surfaces are emphasized. Symmetry is profitably utilized in the fourth example where a symmetric steady state of the well-known Kuramoto-Sivashinsky  equation is investigated whose representation is greatly simplified due to symmetry. In the last example, an exact solution is derived in a $3d$  system with the implicit function approach which may be well extended to other systems.
\subsection*{\textbf{{Application A: Lotka-Volterra model of competition}}}\label{subsec3.1}
\begin{figure}[!htp]
\centering
\subfloat[]{
\begin{minipage}{1.5in}
\centering
\includegraphics[width=1.5in]{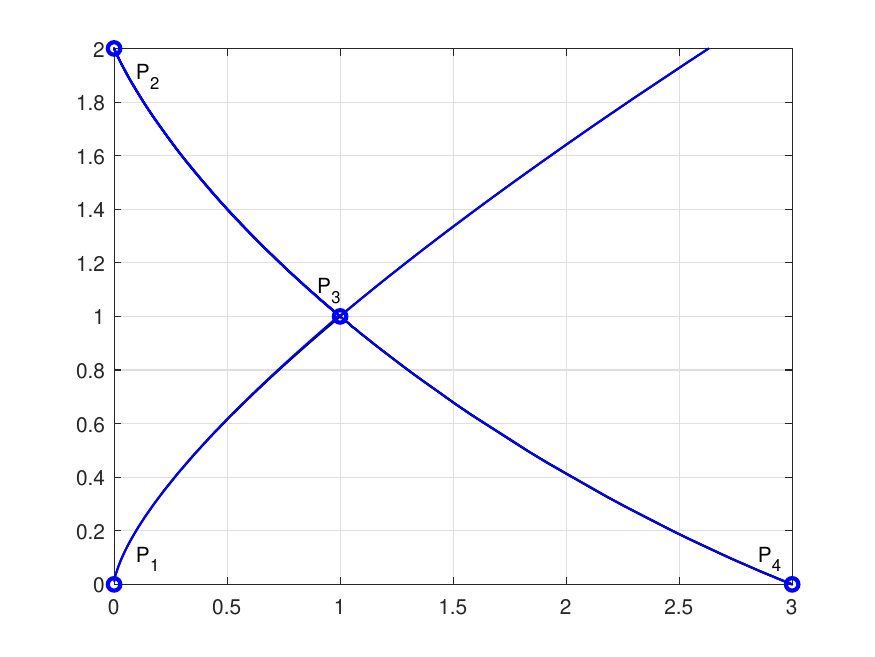}
\end{minipage}
}
\subfloat[]{
\begin{minipage}{1.5in}
\centering
\includegraphics[width=1.5in]{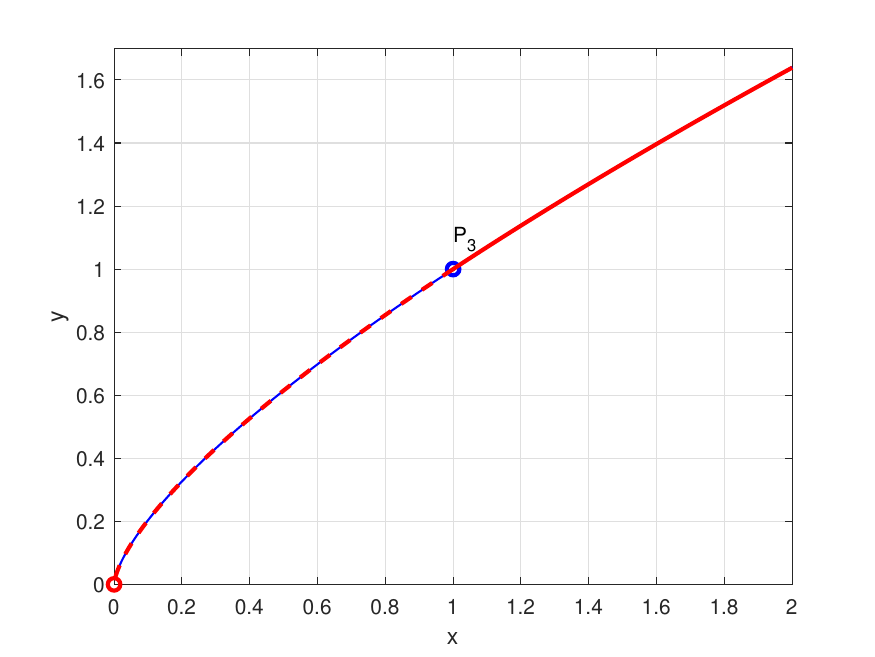}
\end{minipage}
}
\caption{(a).The phase portrait of the Lotka-Volterra model~(\ref{3.1}). (b). The result (red dashed line) from the implicit function (\ref{3.7}) compared with the benchmark computation (blue dotted line). }
\label{Fig1}
\end{figure}

We begin with the classic Lotka-Volterra model of competition as an example to elaborate the effectiveness of the implicit function approach. The model involves two species, here conceived to be rabbits and sheep. They graze on the same lawn and the amount  of grass is limited. When two species meet, conflict between them starts. The following model describes the growth and  the competition between them~\cite{5_2000Nonlinear}
\begin{equation}
	\begin{array}{lll}
\dot{x}&=&x(3-x-2y)\\
\dot{y}&=&y(2-x-y),
	\end{array} \label{3.1}
\end{equation}
where $x$, $y$ are the number of rabbits and sheep, respectively. The cross terms on the right hand side of Eq.~(\ref{3.1}) describes the competition. The phase portrait of this system is shown in Fig.$\ref{Fig1}(a)$. \par
This system has four fixed points:$P_1=(0,0)$, $P_2=(0,2)$, $P_3=(1,1)$, and $P_4=(3,0)$. At the fixed point $P_3$, there are two eigenvalues, $k_1=-1-\sqrt{2}$ and $k_2=-1+\sqrt{2}$, corresponding to the stable and the unstable direction of this point.  We will derive an implicit function based on the RG method to represent the heteroclinic orbit between the fixed points $P_1$ and $P_3$. \par
\par
The new method requires the knowledge of the dynamical behaviors near the two fixed points. Firstly,  we  write the perturbative expansion at point $P_3$
\begin{equation}
	\begin{array}{lll}
x&=&1+\epsilon x_1+\epsilon^2x_2+\epsilon^3x_3+...\\
y&=&1+\epsilon y_1+\epsilon^2y_2+\epsilon^3y_3+....
	\end{array} \label{3.2}
\end{equation}
Inserting $(\ref{3.2})$ into $(\ref{3.1})$ and comparing different orders of $\epsilon$, we have the following linear equations
\begin{subequations}
\begin{align}
\dot{x_1}&=-x_1-2y_1\,,\notag \\
\dot{y_1}&=-x_1-y_1\,, \label{3.3a}\\
\dot{x_2}&=-x_2-2y_2-x_1^2-2x_1y_1\,,\notag\\
\dot{y_2}&=-x_2-y_2-x_1y_1-y_1^2\label{3.3b}\,,\\
&...\notag \,.
\end{align}
\end{subequations}
The general solution of the linear differential Eq. $(\ref{3.3a})$ is $x_1=ce^{k_1(t-t_0)}+de^{k_2(t-t_0)}$, $y_1=\frac{\sqrt{2}}{2}ce^{k_1(t-t_0)}-\frac{\sqrt{2}}{2}de^{k_2(t-t_0)}$. Where $c=c(t_0)$ and $d=d(t_0)$ give the initial condition. We set $d=0$ since only the stable manifold near $P_3$ is what we are concerned about. The solution of $(\ref{3.3a})$ can be simplified to $x_1=ce^{k_1(t-t_0)}$, $y_1=\frac{\sqrt{2}}{2}ce^{k_1(t-t_0)}$, which may be substituted into $(\ref{3.3b})$ and leads to $x_2=\frac{2(5+2\sqrt{2})}{17}c^2e^{2k_1(t-t_0)}$, $y_2=\frac{9+7\sqrt{2}}{34}c^2e^{2k_1(t-t_0)}$. Repeat this process, the expansion of $(\ref{3.1})$ around  $P_3$ can be written as
\begin{equation}
	\begin{array}{lll}
x&=&1+\epsilon c+\epsilon^2\frac{10+4\sqrt{2}}{17}c^2+\epsilon^3\frac{9+10\sqrt{2}}{28}c^3+...\\
y&=&1+\epsilon\frac{c}{\sqrt{2}}+\epsilon^2\frac{9+7\sqrt{2}}{34}c^2+\epsilon^3\frac{290+99\sqrt{2}}{952}c^3+... ,,
	\end{array} \label{3.5}
\end{equation}
after setting $t=t_0$. The RG equation is simply $\frac{dc}{dt_0}=k_1 c$.
\par
The two eigenvalues of the fixed point $P_1$ are $\lambda_1=3$ and $\lambda_2=2$, both unstable, which needs two parameters $a=a(t_0)$ and $b=b(t_0)$ to describe. The approximate solution of $(\ref{3.1})$ near point $P_1$ is then
\begin{equation}
	\begin{array}{lll}
x=&\epsilon a-\epsilon^2\frac{1}{3}a^2-\epsilon^2ab+\epsilon^3\frac{3}{4}ab^2\\
&+\epsilon^3\frac{2}{3}a^2b+\epsilon^3\frac{1}{9}a^3+...\\
y=&\epsilon b-\epsilon^2\frac{1}{2}b^2-\epsilon^2\frac{1}{3}ab+\epsilon^3\frac{1}{4}b^3\\
&+\epsilon^3\frac{13}{30}ab^2+\epsilon^3\frac{1}{9}a^2b+...\,,
\end{array} \label{3.6}
\end{equation}
where $a$ and $b$ satisfy the RG equations $\frac{da}{dt_0}=3a$ and $\frac{db}{dt_0}=2b$ by the standard RG scheme. To derive the relation function for the connection, we may follow the procedure introduced in Section \ref{sec2} and set $a=\beta b^{\frac{3}{2}}$, where  $\beta$ is an unknown  parameter used to select the filament that embeds the heteroclinic orbit. For the first example in our paper, a detailed explanation is provided of how to obtain the relation function and estimate this unknown parameter with the help of the error function. Still along the lines  in Section $\ref{sec2}$,  we derive $a=x+\epsilon(x^2+xy)+...$ and $b=(y+\epsilon(y^2+xy+)...)^{\frac{3}{2}}$ from Eq. $(\ref{3.6})$.  The relation function  contains all the low-order terms in the expansion of the expression $a=\beta b^{\frac{3}{2}}$ treating $x,y$ as small variables
\begin{equation}
	\begin{array}{lll}
f(x,y)&\equiv \epsilon(a_1x+a_2y+a_3y^{\frac{3}{2}})+\epsilon^2(a_4x^2\\
&+a_5y^2+a_6y^3+a_7xy^{\frac{3}{2}}+a_8y^{\frac{5}{2}}+a_9xy)=0,
\end{array} \label{3.7}
\end{equation}
with undetermined coefficients $\{a_{i}\}_{i=1,2,...,9}$. In this computation, we will take $\epsilon=1$.
 Near the fixed point $P_3$, the curve can be described with the parameter $c$, indicating that the implicit function is analytic in $x, y$ there, which is the case for the above $f(x,y)$.  The equality $f(x,y)=0$ then represents a curve in the  $2d$ phase space, which is supposed to capture the analytic behavior near both fixed points $P_1$ and $P_3$.
\par
The coefficients $\{a_{i}\}_{i=1,2,...,9}$ are determined by this analyticity requirement at both ends of the curve. For this, Eq. $(\ref{3.5})$ is first substituted into $f(x,y)=0$. By compare different orders of $c$, we obtain Eqs. $(\ref{3.8a})$ and $(\ref{3.8b})$  corresponding to $c$ and $c^2$ respectively. The point $P_1$ automatically satisfy $f(x,y)=0$ while Eq.~$(\ref{3.8c})$ puts the point $P_3$ on this curve. Eq.~$(\ref{3.8d})-(\ref{3.8h})$ are derived when $(\ref{3.6})$  is substituted into $(\ref{3.7})$ together with the relation $a=\beta b^{\frac{3}{2}}$, by comparing different orders of $b$, {\em i.e.}, $b^{\frac{3}{2}}$, $b^2$, $b^{\frac{5}{2}}$, and $b^3$. In all, the equation system $(\ref{3.8a})-(\ref{3.8h})$ is displayed as follows
\begin{subequations}
\begin{align}
&a_1+(2\sqrt{2}a_2+3\sqrt{2}a_3+8a_4+4\sqrt{2}a_5  \notag\\
&+6\sqrt{2}a_6+(4+3\sqrt{2})a_7+5\sqrt{2}a_8 \notag\\
&+(4+2\sqrt{2}a_9))/4=0; \label{3.8a}\\
&(32(5+2\sqrt{2})a_1+8(9+7\sqrt{2})a_2 \notag\\
&+(159+84\sqrt{2})a_3+(592+128\sqrt{2})a_4 \notag\\
&+(280+112\sqrt{2})a_5+(624+168\sqrt{2})a_6 \notag\\
&+(319+352\sqrt{2})a_7+(435+140\sqrt{2})a_8 \notag\\
&+(232+256\sqrt{2})a_9)/272=0;\label{3.8b} \\
&a_1+a_2+a_3+a_4+a_5+a_6+a_7+a_8+a_9=0;\label{3.8c}\\
&a_2=0;\label{3.8d}\\
&a_3+a_1\beta=0;\label{3.8e}\\
&-a_2/2+a_5=0;\label{3.8f}\\
&-3a_3/4+a_8-(3a_1+a_2-3a_9)/3\beta=0;\label{3.8g}\\
&{a_2}/4-a_5+a_6-{a_3}/2{2}\beta+a_7\beta-{a_1}/3\beta^2+a_4\beta^2=0 \,, \label{3.8h}
\end{align}
\end{subequations}
in which we find that the unknown parameters $\{a_{i}\}_{i=1,2,...,9}$ enter Eqs. $(\ref{3.8a})-(\ref{3.8h})$ linearly, which is one advantage fo the current approach mentioned before. For the same reason, without loss of generality, we may take $a_1=1$ and then all other $a_i$'s could be expressed as a function of $\beta$. The parameter $\beta$ enters the equation system nonlinearly and has to be estimated by minimizing the relative error function $(\ref{new5})$.
\par
For this purpose, we get a numerical representation of the relation function~(\ref{3.7}) according to $(\ref{new4})$ starting with a point $(\tilde{x}_0,\tilde{y}_0)=(10^{-5}, 7\times 10^{-6})$ in the unstable direction at an initial guess $\beta=1.05$. The sampling time is $(t=0:0.01:7.6)$ so that the curve is depicted with a set of $761$ data points $(x_j,y_j)_{j=0:1:760}$. With some arbitrariness, from this data set, $16$ points are uniformly selected starting from the $50th$ one with a stepsize of $50$. Subsequently, the unknown parameter $\beta$ is estimated by minimizing the error function $\Delta(\beta)$ in Eq.~(\ref{new5}). We get $\beta=1.1317$ at which the relative error  $\Delta(\beta)=1.49\times 10^{-11}$ and the values of the coefficients $\{a_{i}\}_{i=1,2,...,9}$ are shown in Table $\ref{T1}$. In Fig.$\ref{Fig1}(b)$, we compare the orbit from the implicit  function (red dashed line) and from the benchmark numerical computation (blue dotted line).\par
\begin{table}[h]
\centering
\begin{tabular}{@{}llll@{}}
\toprule
$a_{1}=1$  &  $a_{2}=0$  & $a_{3}=-1.1317$ \\
$a_{4}=-0.2352$&  $a_{5}=0$  & $a_{6}=-0.0855$ \\
$a_{7}=0.1532$&  $a_{8}=0.4235$  & $a_{9}=-0.1242$  \\
\bottomrule
\end{tabular}
\caption{The values of  coefficients $\{a_{i}\}_{i=1,2,...,9}$  in the relation function $(\ref{3.7})$.}\label{T1}%
\end{table}
Interestingly, the implicit function $(\ref{3.7})$ not only represents the trajectory from point $P_1$ to $P_3$, but also the trajectory beyond point $P_3$ shown as the solid red line in Fig.$\ref{Fig1}(b)$. So, it is a geometric representation of invariant curves of the system~(\ref{3.1}) compatible with the boundary condition at $P_1$ and $P_3$. In this example, the relation function is a generalized polynomial with non-integer powers to account for the presence of two unstable directions at the starting point of the heteroclinic orbit.

\subsection*{\textbf{{Application B: System with Homoclinic Bifurcation}}}\label{subsec3.2}\par
In this section, the framework will be used to search for orbits with spiral structure or a homoclinic orbit in
 a  $2d$  system appearing in Strogatz's book~\cite{5_2000Nonlinear}
\begin{equation}
	\begin{array}{lll}
\dot{x}&=&y\\
\dot{y}&=&\mu y+x-x^2+xy,
	\end{array} \label{3.10}
\end{equation}
where $x$ and $y$ are the state variables and $\mu$ is an unknown parameter. Two fixed points  $P_1=(0,0)$ and $P_2=(1,0)$  exist. This system experiences interesting bifurcations as $\mu$ changes. From the phase portraits~\cite{5_2000Nonlinear}, a homoclinic orbit should exist at a specific value $\mu = \mu_c$ within the interval $[-0.9, -0.8 ]$.  When $\mu>\mu_c$, we have a heteroclinic orbit connecting $P_2$ and $P_1$.   In the following, we will approximate both connections and estimate the critical value $\mu_c$.

\subsubsection*{\textbf{{Case1. Implicit function for homoclinic orbit}}}\label{subsec3.2.1}
 First, we work out a relation function that describes the homoclinic orbit. At the origin, The two eigenvalues $k_1=\frac{1}{2}(\mu-\sqrt{4+\mu^2})$, $k_2=\frac{1}{2}(\mu+\sqrt{4+\mu^2})$ determine. local dynamics with $k_1<0$, $k_2>0$. The expansion near the origin of Eq. $(\ref{3.10})$ can  be expressed in a polynomial of $a=a(t_0)$ and $b=b(t_0)$, where the renormalized constants  $a,b$ denote the initial positions corresponding to the eigenvalues $k_1$ and $k_2$ respectively. For the expansion along the stable (unstable) direction, we take $b(t_0)=0(a(t_0)=0)$, which gives
\begin{equation}
	\begin{array}{lll}
x&=&\epsilon a+\epsilon^2q_1a^2+\epsilon^3q_2a^3...\\
y&=&\epsilon k_1 a+\epsilon^22k_1q_1a^2+\epsilon^33k_1q_2a^3...,
\end{array} \label{3.11}
\end{equation}
and
\begin{equation}
	\begin{array}{lll}
x&=&\epsilon b+\epsilon^2p_1b^2+\epsilon^3p_2b^3...\\
y&=&\epsilon k_2 b+\epsilon^22k_2p_1b^2+\epsilon^33k_2p_2b^3...\,,
\end{array} \label{3.12}
\end{equation}
where the coefficients $\{p_i\}$ and $\{q_{i}\}_{i=1,2,...n}$ can be written as a function of $\mu$, which  are displayed in $Appendix~~A(1)$. The relation function is given by a polynomial
\begin{equation}
	\begin{array}{lll}
f(x,y)&=\epsilon(a_1x+a_2y)+\epsilon^2(a_3x^2+a_4y^2+a_5xy)\\
&+\epsilon^3(a_6x^3+a_7y^3+a_8x^2y+a_9xy^2)+\epsilon^4(a_{10}x^4\\
&+a_{11}y^4+a_{12}x^3y+a_{13}xy^3+a_{14}x^2y^2)=0
\end{array} \label{3.13}
\end{equation}
since the stable and the unstable manifold is just  $1d$  on both ends.  Again, we set $\epsilon=1$.  Substituting $(\ref{3.11})$ or $(\ref{3.12})$ into $(\ref{3.13})$ and comparing the different orders of $\{a^{k}\}_{k=1,2,...,7}$ or $\{b^{j}\}_{j=1,2,...,6}$, we obtain linear  equations similar to Eqs. $(\ref{3.8a})-(\ref{3.8h})$. Due to space limitations, we will not provide the specific expressions for such linear equations started from this example.\par
\begin{figure}[!htp]
\centering
\subfloat[]{
\begin{minipage}{1.5in}
\centering
\includegraphics[width=1.5in]{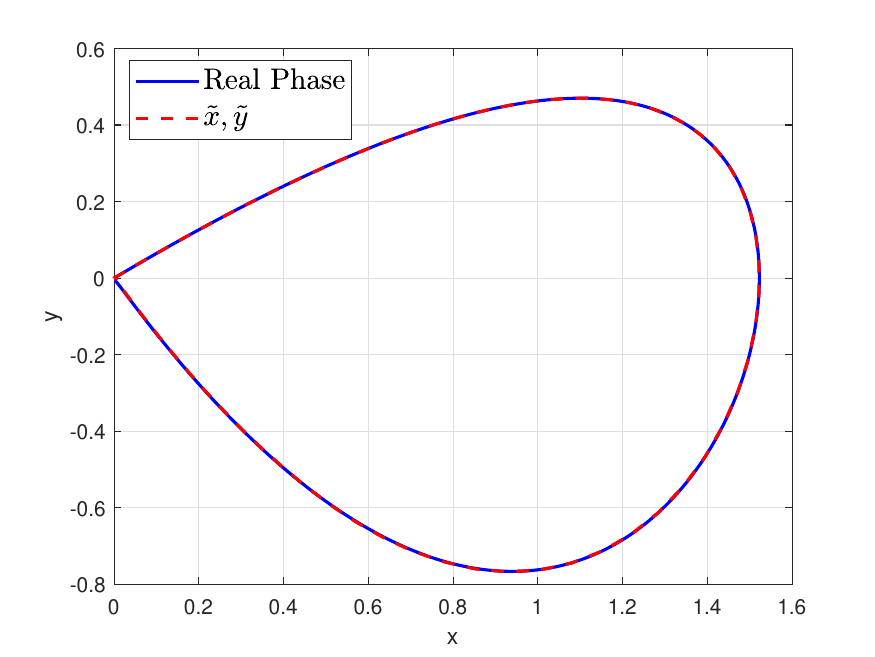}
\end{minipage}
}
\subfloat[]{
\begin{minipage}{1.5in}
\centering
\includegraphics[width=1.5in]{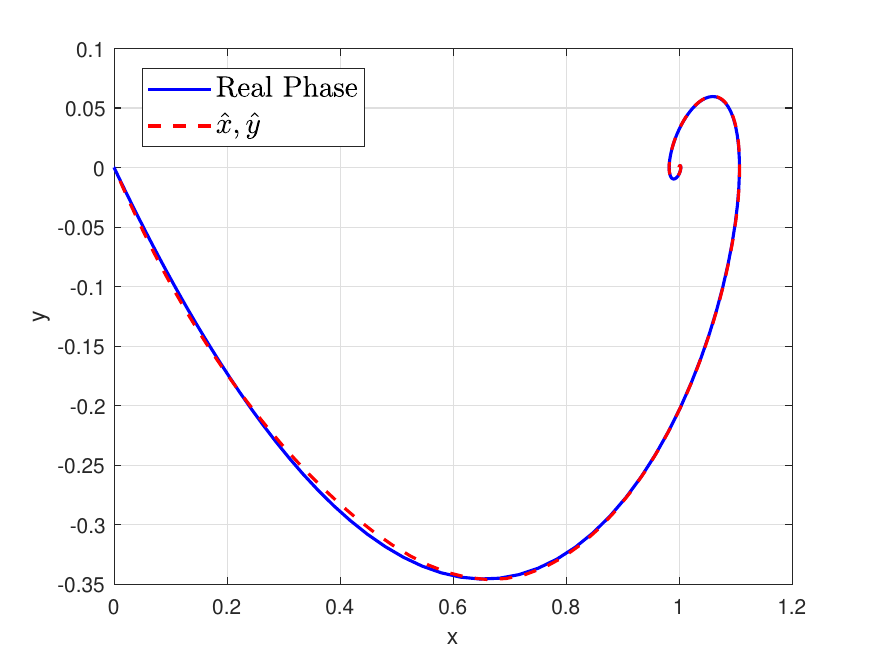}
\end{minipage}
}
\caption{The connecting orbits in the $2d$  system~(\ref{3.10}). (a) The homoclinic connection and (b) a spiraling connection determined by the relation function~(\ref{3.13}) (red dotted line) and the numerical benchmark (blue solid line).  }
\label{Fig3}
\end{figure}

  With the same procedure as detailed in section $\ref{sec2}$,  we obtain a trial orbit starting with the initial position $(\tilde{x}_0,\tilde{y}_0)=(0.8367\times 10^{-5},0.5492\times 10^{-5})$ and the parameter $\mu_0=-0.8$. In this case, the orbit is depicted with $251$ sampling points and different from what is done in  $Application~A$, only $4$ points ($193 - 196$) are selected to the error function $(\ref{new5})$.   After the coefficient $a_3=1$ is taken, a minimization of the error function delivers $\mu_c=-0.8644$ with $\Delta(\mu)=2.93\times 10^{-9}$ and thus the coefficients $\{a_{i}\}_{i=1,2,...,14}$ as displayed in Table $\ref{tT2}$.   The corresponding homoclinic orbit is shown in Fig.$\ref{Fig3}(a)$ as the red dotted curve, which just overlaps with the benchmark numerical solution (blue solid line).\par
 In Strogatz's book~\cite{5_2000Nonlinear} $\tilde{\mu}_c=-0.8645$, which is very close to our result.  Considering the low-order approximation, it is easy to see that the implicit function in the current design  offers a very accurate approximation. Besides, the determination of the unknown parameter such as $\mu$ is quite flexible, as long as the error function $\Delta(\mu)$ is made small.  \par
\begin{table}[h]
\centering
\begin{tabular}{@{}llll@{}}
\toprule
$a_{1}=0$  &  $a_{2}=0$  & $a_{3}=1$ & $a_4=-1$  \\
$a_5=-0.8644$ & $a_{6}=-0.78$ &  $a_{7}=-0.003$  & $a_{8}=0.6917$ \\
 $a_{9}=0.1456$  & $a_{10}=0.0809$ & $a_{11}=0$ &  $a_{12}=-0.0813$  \\
 $a_{13}=0$ &  $a_{14}=0$ \\
 \bottomrule
\end{tabular}
\caption{The values of  coefficients $a_{i,i=1,2,...,14}$ in the relation function $(\ref{3.13})$.}\label{tT2}%
\end{table}
\par
\subsubsection*{\textbf{{{Case2. Implicit function for a heteroclinic orbit with single spiral structure}}}}\label{subsec3.2.2}
 \
 \newline
As we have claimed before,  when $\mu> \mu_c$, a heteroclinic orbit with a single spiral structure is present. We take $\mu=0$ for convenience and the two eigenvalues $k_1=-1$ and $k_2=1$ now. At the fixed point $P_2=(1,0)$, there exist a pair of conjugate eigenvalues $\lambda_{1}=\frac{1+i\sqrt{3}}{2}$, $\lambda_2=\frac{1-i\sqrt{3}}{2}$ with positive real parts which explains the expanding spiraling structure around this point. Thus, the connection matches the stable direction at the origin. Before providing the approximate expansion, for convenience,  we introduce the coordinate transformation
\begin{equation}
\begin{array}{lll}
u=\frac{3+i\sqrt{3}}{6}(x-1)-\frac{i}{\sqrt{3}}y\\
v=\frac{3-i\sqrt{3}}{6}(x-1)+\frac{i}{\sqrt{3}}y\,.
\end{array} \label{3.150}
\end{equation}
The perturbative expansion along the stable direction around the origin is
\begin{equation}
\begin{array}{lll}
u&=&\frac{-3-i\sqrt{3}}{6}(1+\epsilon \frac{-3-i\sqrt{3}}{2}c+\epsilon^2\frac{4+2i\sqrt{3}}{3}c^2...)\\
v&=&\frac{-3+i\sqrt{3}}{6}(1+\epsilon \frac{-3+i\sqrt{3}}{2}c+\epsilon^2\frac{4-2i\sqrt{3}}{3}c^2...),
\end{array} \label{3.15}
\end{equation}
and that near the fixed point $P_2$ is
\begin{equation}
\begin{array}{lll}
u&=&\epsilon {a}+\epsilon^2(\frac{3-i\sqrt{3}}{6}{a}^2+\frac{-3+i\sqrt{3}}{6}{a}{b}-\frac{3+2i\sqrt{3}}{21}{b}^2)+...\\
v&=&\epsilon {b}+\epsilon^2(\frac{3+i\sqrt{3}}{6}{b}^2+\frac{-3-i\sqrt{3}}{6}{a}{b}-\frac{3-2i\sqrt{3}}{21}{a}^2)+...,
\end{array} \label{3.16}
\end{equation}
where $c$, ${a}$, and ${b}$ satisfy $\frac{dc}{dt_0}=k_1c$, $\frac{d{a}}{dt_0}=\lambda_1{a}$, and $\frac{d{b}}{dt_0}=\lambda_2{b}$ based on the RG equation. Thus, in polar coordinates, we may write ${a}=d^{\alpha}e^{i\theta}$ and ${b}=d^{\bar{\alpha}}e^{-i\theta}$, where  $\alpha=21+i\sqrt{3}$ and $\bar{\alpha}$ is the conjugate of $\alpha$. The variable $d$  is the growing radius of the spiral and $\theta$ is the initial phase. In this subsection, we take $\theta=5.320$ which could be fixed with the error function $(\ref{new5})$ but a straightforward alternative way to determine $\theta$ is presented in $Appendix~A(2)$.  As before, the equality ${a}^{\bar{\alpha}}e^{-i\bar{\alpha}\theta}={b}^{\alpha}e^{i\alpha\theta}$ combined with the inverse of Eq.~(\ref{3.16}) ${a}=\epsilon u+\epsilon^2(u^2+v^2+uv)+...$, ${b}=\epsilon v+\epsilon^2(v^2+u^2+uv)+...$ suggests the form of the relation function truncated at $\epsilon^2$
 \begin{equation}
\begin{array}{lll}
\hat{f}(u,v)&=\epsilon e^{-i\bar{\alpha}\theta}u^{\bar{\alpha}}(r_1+\epsilon(r_2u+r_3v+r_4\frac{v^2}{u}))\\
&+\epsilon e^{i\alpha\theta}v^{\alpha}(\bar{r}_1+\epsilon(\bar{r}_2v+\bar{r}_3u+\bar{r}_4\frac{u^2}{v}))=0,
\end{array} \label{3.17}
\end{equation}
 where $\{r_{j}\}_{j=1,2,3,4}$  and its conjugate  $\bar{r}_j$ are the coefficients. We require that the relation function approximate  well the spiral structure near point $P_2$ and the stable submanifold near the origin simultaneously. Thus we substitute $(\ref{3.15})$ or $(\ref{3.16})$ with ${a}=d^{\alpha}e^{i\theta}$, ${b}=d^{\bar{\alpha}}e^{-i\theta}$  into $(\ref{3.17})$ respectively. Two sets of linear equations are obtained by  comparing different powers of $c$ and $d$: $c$, $d^{\alpha\bar{\alpha}}$, $d^{\alpha(\bar{\alpha}+1)}$, $d^{\bar{\alpha}(\alpha+1)}$, $d^{2\alpha-\bar{\alpha}+\alpha\bar{\alpha}}$ and $d^{2\bar{\alpha}-\alpha+\alpha\bar{\alpha}}$. Besides, the origin should satisfy Eq. $(\ref{3.17})$, giving another equation and all together we have seven equations now.  Finally,  by setting $r_1=i$, we solve these equations for the coefficients $\{r_j,\bar{r}_j\}_{j=1,2,3,4}$ which are displayed in Table $\ref{T3}$. The relation function of $x,y$ is obtained by inserting $(\ref{3.150})$ into $(\ref{3.17})$, which yields the orbit (red dotted lin) in Fig.$\ref{Fig3}(b)$, well matching the numerical one (blue solid line).\par
\begin{table}[h]
\centering
\begin{tabular}{@{}ll@{}}
\toprule
$r_{1}=i$  & $\bar{r}_{1}=-i$  \\
$r_{2}=0.5985+0.4103i$&  $\bar{r}_{2}=0.5985-0.4103i$ \\
$r_{3}=-0.5985+0.4103i$  &$\bar{r}_{3}=-0.5985-0.4103i$ \\
$r_{4}=0.0825+0.4286i$ &  $\bar{r}_{4}=0.0825-0.4286i$\\
\bottomrule
\end{tabular}
\caption{The values of  coefficients $\{r_j,\bar{r}_j\}_{j=1,2,3,4}$ in relation function $(\ref{3.17})$.}\label{T3}%
\end{table}
In applications $A$ and $B$, we have proven that the implicit function based on the RG method may be used to approximate heteroclinic orbit, homoclinic orbit, and heteroclinic orbit with spiral structure in  $2d$  dynamical systems with good performance. Next, this technique will be applied to  $3d$  dynamical systems.

\subsection*{\textbf{{{Application C: Lorenz System}}}}\label{subsec3.3}
In this application, we study the homoclinic and heteroclinic orbit in the famous Lorenz system. In paper \cite{9_1963Deterministic},  Lorenz derived this $3d$ system from a greatly simplified model depicting convection rolls in the atmosphere.  This model serves a critical role in the study of key phenomena such as chaos, periodic orbits, and various bifurcations\cite{41_1982The, 42_1993New,43_Konstantin1995Chaos}. Here, our primary focus is on approximating connecting orbits with implicit functions. The Lorenz equation is
 \begin{equation}
\begin{array}{lll}
\dot{x}&=&\sigma(y-x)\\
\dot{y}&=&rx-y-xz\\
\dot{z}&=&xy-\beta z,
\end{array} \label{3.18}
\end{equation}
where $x$, $y$, and $z$ are the state variables with there parameters $\sigma$,  $\beta$, and the Rayleigh number $r$. To explore the bifurcation, many researchers opt to maintain constant values for $\sigma=10$ and $\beta=\frac{8}{3}$ while varying $r$. In this convention,  a homoclinic orbit emerges at the bifurcation point of  $r=\tilde{r}$.  When $1<r<\tilde{r}$, a heteroclinic orbit  with a spiral structure continue to exist.  For convenience, we take $r=5$ and check this heteroclinic orbit with the current scheme.

\subsubsection*{\textbf{{Case1. Implicit function for a heteroclinic orbit with single spiral structure for Lorenz system}}}\label{subsec3.3.1}
When $r=5$, there exists a heteroclinic orbit connecting the origin $(0,0,0)$ and the fixed point $M=(4\sqrt{\frac{2}{3}},4\sqrt{\frac{2}{3}},4)$. At the point $M$, the stability matrix of Eq.(\ref{3.18}) has three eigenvalues $k_1=-0.541+6.487i$, $k_2=-0.541-6.487i$, and $k_3=-12.584$.  To simplify the computation, we make the following coordinate transformation
\begin{equation}
\begin{array}{lll}
u=g_{10}+g_{11}x+g_{12}y+g_{13}z\\
v=g_{20}+g_{21}x+g_{22}y+g_{23}z\\
w=g_{30}+g_{31}x+g_{32}y+g_{33}z
\end{array} \label{3.19}
\end{equation}
to diagonalize the stability matrix, where $\{g_{ij}\}_{i,j=1,2,3}$ are the corresponding diagonaling matrix elements displayed in Table $\ref{AT1}$ of  $Appendix~~B(1)$. If we only consider the submanifold that embeds the spiral, the expansion is
\begin{equation}
\begin{array}{lll}
u=\epsilon a+\epsilon^2(p_{11}a^2+p_{21}b^2+p_{31}ab)+...\\
v=\epsilon b+\epsilon^2(p_{12}a^2+p_{22}b^2+p_{32}ab)+...\\
w=\epsilon^2(p_{13}a^2+p_{23}b^2+p_{33}ab)+...
\end{array} \label{3.20}
\end{equation}
to the second order of $\epsilon$ where the values of the parameters $\{p_{ij}\}_{i,j=1,2,3}$ are shown in Table $\ref{AT2}$ of $Appendix~~B(1)$. The expansion at the origin involves only the unstable direction which is parameterized by $c$
\begin{equation}
\begin{array}{lll}
u=u_{1}+\epsilon u_{2}c+\epsilon^2u_{3}c^2...\\
v=v_{1}+\epsilon v_{2}c+\epsilon^2v_{3}c^2...\\
w=w_{1}+\epsilon w_{2}c+\epsilon^2w_{3}c^2...,
\end{array} \label{3.21}
\end{equation}
where $\{u_j,v_j,w_{j}\}_{j=1,2,3}$ are placed in Table $\ref{AT3}$ of $Appendix~B(1)$. With similar procedures  in Section $\ref{sec2}$, we write $a=d^{\alpha}e^{i\theta}$ and $b=d^{\bar{\alpha}}e^{-i\theta}$, where $\alpha=1-4.466i$ and $\bar{\alpha}=1+4.466i$, and acquire the following relation function to denote the spiral structure
\begin{equation}
\begin{array}{lll}
f_1(u,v)&=\epsilon u^{\bar{\alpha}}e^{-i\bar{\alpha}\theta}(a_1+\epsilon(a_2u+a_{3}\frac{v^2}{u}+a_4v))\\
&+\epsilon v^{\alpha}e^{i\alpha\theta}(\bar{a}_1+\epsilon(\bar{a}_2v+\bar{a}_{3}\frac{u^2}{v}+\bar{a}_4u))=0,\\
\end{array} \label{3.22}
\end{equation}
where $\{a_j,\bar{a}_{j}\}_{j=1,2,3,4}$ are coefficients to be determined and $\theta=2.012$ can be estimated with the error function or the same as in $Application~B$. After substituting the expansions $(\ref{3.20})$ or $(\ref{3.21})$ into the relation function (\ref{3.22}) and comparing different orders of $\epsilon$, as before, we have a set of linear equations of  $\{a_j,\bar{a}_{j}\}_{j=1,2,3,4}$, the values of which are then obtained and displayed in Table $\ref{T4}$.
\begin{table}[h]
\centering
\begin{tabular}{@{}ll@{}}
\toprule
$a_{1}=i$  & $\bar{a}_{1}=-i$  \\
$a_{2}=-0.419-0.201i$&  $\bar{a}_{2}=-0.419+0.201i$ \\
$a_{3}=0.015+0.108i$  &$\bar{a}_{3}=0.015-0.108i$ \\
 $a_{4}=0.418+0.015i$ &  $\bar{a}_{4}=0.418-0.015i$\\
 \bottomrule
\end{tabular}
\caption{The values of the coefficients $\{a_j,\bar{a}_{j}\}_{j=1,2,3,4}$ in the relation function $(\ref{3.22})$.}\label{T4}%
\end{table}\par
As shown in equations $(\ref{3.7})$, $(\ref{3.13})$, and $(\ref{3.17})$, one implicit function represent a curve in a $2d$ phase space.  Here, the implicit function depicts a spiraling surface in the $3d$ phase space since Eq.~$(\ref{3.22})$ involves complex powers. In order to obtain a curve in $3d$ phase space, we need another surface and the intersection of these two surfaces is the curve we need. The second surface that embeds the spiraling structure has a polynomial form
\begin{equation}
\begin{array}{lll}
f_2(x,y,z)&\equiv \epsilon(c_1x+c_2y+c_3z)+\epsilon^2(c_4x^2+c_5y^2\\
&+c_6z^2+c_7xy+c_8xz+c_9yz)+\epsilon^3(c_{10}x^3\\
&+c_{11}y^3+c_{12}z^3+c_{13}x^2y+c_{14}x^2z\\
&+c_{15}y^2z+c_{16}xz^2+c_{17}yz^2\\
&+c_{18}xy^2+c_{19}xyz)=0,
\end{array} \label{3.25}
\end{equation}
where $\{c_{j}\}_{j=1,2,...,19}$ are the undetermined coefficients. In contrast to the above treatment, substituting Eq.~(\ref{3.20}) (together with (\ref{3.19})) into (\ref{3.25}), we treat  $a$ and $b$ as independent variables, parameterizing  the surface where all spirals lie. The substitution at the other end - Eq.~(\ref{3.21}) to (\ref{3.25}) is the same as before. With this considerations,  we are able to obtain the values of $\{c_{j}\}_{j=1,2,...,19}$, listed in Table $\ref{T5}$ after taking $\epsilon=1$.
\begin{table}[h]
\centering
\begin{tabular}{@{}llll@{}}
\toprule
$c_{1}=4.079$  & $c_{2}=-3.167$ &$c_3=0.015$& $c_4=1$ \\
$c_5=0.357$  &$c_{6}=0.037$  & $c_{7}=-1.237$ &$c_8=-0.874$  \\
$c_9=0.483$&$c_{10}=0.531$ &$c_{11}=-0.280$  & $c_{12}=-0.022$   \\
$c_{13}=-1.229$& $c_{14}=-0.437$&$c_{15}=-0.262$&$c_{16}=0.156$  \\
 $c_{17}=-0.113$ &$c_{18}=1.005$& $c_{19}=0.661$\\
 \bottomrule
\end{tabular}
\caption{The values of  coefficients $\{c_{j}\}_{j=1,2,...,19}$ in the relation function $(\ref{3.25})$.}\label{T5}%
\end{table}\par
\begin{figure}[hbtp]
\centering
\subfloat[]{
\begin{minipage}{1.5in}
\centering
\includegraphics[width=1.5in]{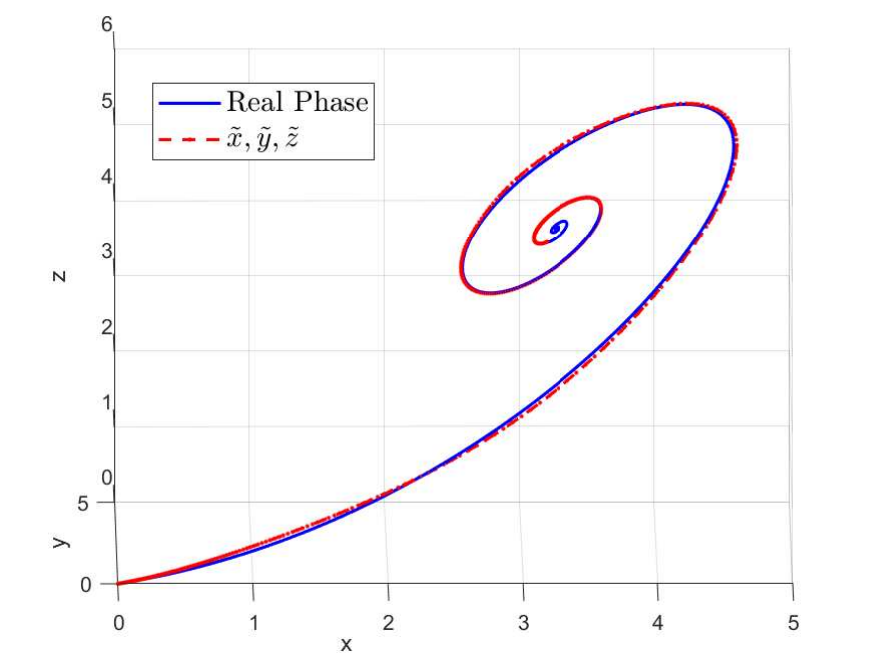}
\end{minipage}
}
\subfloat[]{
\begin{minipage}{1.5in}
\centering
\includegraphics[width=1.5in]{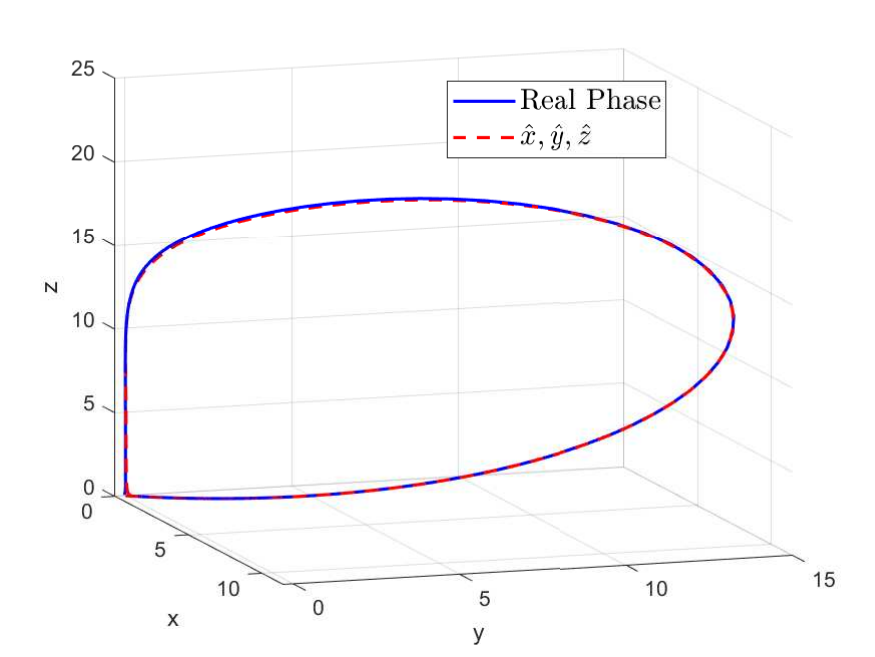}
\end{minipage}
}
\caption{Connections in the Lorenz system~(\ref{3.18}) at different parameter values. A spiraling heteroclinic orbit (a) and the homoclinic orbit (b) computed with the relation function (red dashed line) and from the numerical benchmark (blue solid line). }
\label{Fig5}
\end{figure}

In this application, the function $f_1(u,v)$ is obtained in a similar way to previous cases with a spiraling structure, while a second surface $f_2(x,y,z)=0$ is needed which is a slightly curved surface that contains the spirals. The intersection of these two surfaces is the heteroclinic orbit, which could be obtained by evolving the differential equation given in $(\ref{new4})$ derived from implicit functions  $(\ref{3.25})$.  The result is shown in Fig.$\ref{Fig5}(a)$ with red dashed spiraling line, being compared well to the numerical benchmark (blue solid line).
\par
This design of embedding the connection to different surfaces works well in general. As an example, next we  show how to choose two surfaces to embed the homoclinic orbit in the Lorenz system $(\ref{3.18})$.

\subsubsection*{\textbf{{Case2. Implicit function for homoclinic orbit in the Lorenz system}}}\label{subsec3.3.2}
In this subsection, we aim to obtain an approximate analytical solution for the homoclinic orbit together with the correct value $\tilde{r}$ of the parameter $r$.  It is easy to find that the origin $(0,0,0)$ is a saddle point with one unstable direction and two stable direction when $r\rightarrow \tilde{r}$, since the stability matrix has three eigenvalues $\lambda_1=\frac{-\sqrt{81+40r}-11}{2}$, $\lambda_2=-\frac{8}{3}$, and $\lambda_3=\frac{\sqrt{81+40r}-11}{2}$ at the origin.  Similar to the practice above, we make the following diagonalizing coordinate transformation
\begin{equation}
\begin{array}{lll}
u&=\frac{-(10+\lambda_3)x+10y}{\lambda_1-\lambda_3}\\
v&=z\\
w&=\frac{(10+\lambda_3)x-10y}{\lambda_1-\lambda_3}.
\end{array} \label{3.27}
\end{equation}\par
The expansion at the origin along the unstable direction gives
\begin{equation}
\begin{array}{lll}
u&=\epsilon^2 n_{11}c^3+\epsilon^4 n_{12}c^5+\epsilon^6n_{13}c^7+...\\
v&=\epsilon n_{21}c^2+\epsilon^3 n_{22}c^4+\epsilon^5n_{23}c^6+...\\
w&=c+ \epsilon^2n_{31}c^3+\epsilon^4 n_{32}c^5+\epsilon^6n_{33}c^7+...,
\end{array} \label{3.28}
\end{equation}
where $\{n_{jk}\}_{j,k=1,2,3}$ are the coefficients as listed in Eq. $(\ref{app2})$ of $Appendix~~B(2)$. The expansion along the stable submanifold is
\begin{equation}
\begin{array}{lll}
u&={a}+\epsilon m_{11}{a}{b}+\epsilon^2(m_{12}{a}{b}^2+m_{13}{a}^3)+...\\
v&={b}+\epsilon m_{21}{a}^2+\epsilon^2m_{22}{a}^2{b}+...\\
w&=\epsilon m_{31}{a}{b}+\epsilon^2(m_{32}{a}{b}^2+m_{33}{a}^3)+...\,,
\end{array} \label{3.29}
\end{equation}\\
where the coefficients $\{m_{jk}\}_{j,k=1,2,...}$ are listed in Eq. $(\ref{app3})$ of $Appendix~~B(2)$. The connection should lie in this $2d$ stable manifold. Thus, we may treat $a,b$ as independent and propose the following polynomial function
\begin{equation}
\begin{array}{lll}
F_{1}(u,v,w)&\equiv \epsilon({a}_1u+{a}_2v+{a}_3w)+\epsilon^2({a}_4u^2\\
&+{a}_5uv+{a}_6v^2+{a}_7uw+{a}_8vw\\
&+{a}_9w^2)+\epsilon^3({a}_{10}u^3+{a}_{11}u^2v+{a}_{12}uv^2\\
&+{a}_{13}v^3+{a}_{14}u^2w+{a}_{15}uvw+{a}_{16}v^2w\\
&+{a}_{17}uw^2+{a}_{18}vw^2+{a}_{19}w^3)+\epsilon^4({a}_{20}u^4\\
&+{a}_{21}u^3v+{a}_{22}u^2v^2+{a}_{23}uv^3+{a}_{24}v^4\\
&+{a}_{25}u^3w+{a}_{26}u^2wv+{a}_{27}uv^2w\\
&+{a}_{28}v^3w+{a}_{29}u^2w^2+{a}_{30}uvw^2\\
&+{a}_{31}v^2w^2+{a}_{32}uw^3+{a}_{33}vw^2\\
&+{a}_{34}w^4)=0
\end{array} \label{3.30}
\end{equation}\\
to embed the connecting orbit, where $\{a_{j}\}_{j=1,2,...,34}$ are the undetermined coefficients. For better approximation, here the relation function is taken to  the fourth order.  After the substitution of Eqs. $(\ref{3.28})$ and $(\ref{3.29})$ to Eq. $(\ref{3.30})$, a comparison of different orders - ${a}$, ${b}$, ${a}^2$, ${b}^2$, $...$, ${a}^6$, ${b}^6$, ${a}^5{b}$, ${a}{b}^5$, ${a}^4{b}^2$, ${a}^2{b}^4$,  ${a}^3{b}^3$,  and $\{c^{k}\}_{k=1,2,...,5}$ leads to a system of linear equations which can be solve for the coefficients $\{a_{j}\}_{j=1,2,...34}$ as a function of the parameter $r$.
\par
Above, we provided an approximation of the heteroclinic orbit at $r=5$.  This orbit continues to exist with the increase of $r$ but it gets closer and closer to the origin until finally at $r=\tilde{r}$ it returns exactly making up a homoclinic orbit. Therefore, it is reasonable to choose the surface containing the spirals around the fixed point $M_{1}=(x_0,y_0,z_0)=(2\sqrt{\frac{2}{3}(r-1)},2\sqrt{\frac{2}{3}(r-1)},r-1)$. The stability matrix
\begin{equation}
A=
\left [
\begin{array}{lll}
-10 & 10 & 0 \\
1 & -1 & -x_0 \\
y_0 & x_0 & -\frac{8}{3} \\
\end{array}
\right ]\label{3.31}
\end{equation}
has three eigenvalues $k_1$,  $k_2$, and $k_3$, where $k_1$,  $k_2$ are conjugates with negative real parts and  $k_3<0$.  As before, we carry out an expansion at $M_1$ along the spiraling direction with variables $a, b$ and at the origin along the unstable direction with variable $d$. The details are relegated to $Appendix$ \ref{secAppB2}. As $a, b$ is treated as independent, it is not hard to write a polynomial to characterize the second surface
\begin{equation}
\begin{array}{lll}
F_{2}(x,y,z)&=\epsilon({b}_1x+{b}_2y+{b}_3z)+\epsilon^2({b}_4x^2+{b}_5y^2\\
&+{b}_6z^2+{b}_7xy+{b}_8xz+{b}_9yz)+\epsilon^3({b}_{10}x^3\\
&+{b}_{11}y^3+{b}_{12}z^3+{b}_{13}x^2y+{b}_{14}x^2z\\
&+{b}_{15}y^2z+{b}_{16}xz^2+{b}_{17}yz^2+{b}_{18}xy^2\\
&+{b}_{19}xyz)=0,
\end{array} \label{3.34}
\end{equation}\\
where $\{b_{j}\}_{j=1,2,...19}$ are the coefficients to be determined by the substitution and comparison procedure. For reference, the linear equations  are obtained by comparing the coefficients of $\{d^k\}_{k=1,2,3,4}$,  ${a}$, ${b}$, ${a}{b}$, ${a}^2$, ${b}^2$, $...$, ${a}^4$, ${b}^4$, ${a^3}{b}$, ${a}{b}^3$, and ${a}^2{b}^2$.
\par
The two linear equation systems deduced from relation functions  $F_{1}$ and $F_{2}$ both depend on the system parameter $r$. An estimation of  the $r$ is achieved by minimizing the error function $(\ref{new5})$, which results in $\hat{r}=13.9374$, very close to the benchmark value $\tilde{r}=13.92653$~\cite{5_2000Nonlinear}.  The coefficients $\{a_{j},{b}_{j}\}_{j=1,2,...}$ are also computed and listed in Table $\ref{T6}$ and Table $\ref{T7}$. In Fig.$\ref{Fig5}(b)$, the blue solid line plots the benchmark orbit, in good comparison of  the intersection (red dashed line) of two surfaces described by $(\ref{3.30})$ and $(\ref{3.34})$.
\begin{table}[h]
\centering
\begin{tabular}{@{}llll@{}}
\toprule
${a}_{1}=0$  & ${a}_{2}=0$ &${a}_3=0$& ${a}_4=0$ \\
${a}_5=0$ & ${a}_{6}=0$  & ${a}_{7}=1$ &${a}_8=0$   \\
${a}_9=0$&${a}_{10}=0$&${a}_{11}=-0.014$  & ${a}_{12}=0$  \\
${a}_{13}=0$& ${a}_{14}=0$&${a}_{15}=-0.032$ &${a}_{16}=0$    \\
${a}_{17}=0$ &${a}_{18}=5.720$& ${a}_{19}=0$&${a}_{20}=0$ \\
${a}_{21}=0$  & ${a}_{22}=0$ &${a}_{23}=0$& ${a}_{24}=0$ \\
${a}_{25}=0$ &${a}_{26}=0$  & ${a}_{27}=-0.084$ &${a}_{28}=0$  \\
${a}_{29}=0.086$&${a}_{30}=0$ &${a}_{31}=-0.165$  & ${a}_{32}=0.039$ \\
${a}_{33}=0$& ${a}_{34}=-0.580$\\
 \bottomrule
\end{tabular}
\caption{The values of  coefficients $\{a_{j}\}_{j=1,2,...34}$ in the relation function $(\ref{3.30})$.}\label{T6}%
\end{table}
\begin{table}[h]
\centering
\begin{tabular}{@{}llll@{}}
\toprule
${b}_{1}=-1.713$  & ${b}_{2}=1$ &${b}_3=0.053$& ${b}_4=0.837$  \\
${b}_5=0.193$&${b}_{6}=-0.021$  & ${b}_{7}=-0.822$ &${b}_8=-0.354$  \\
${b}_9=0.235$ &${b}_{10}=0.131$&${b}_{11}=-0.016$  & ${b}_{12}=-0.002$   \\
$b_{13}=-0.207$& ${b}_{14}=-0.142$&${b}_{15}=-0.028$&${b}_{16}=0.035$ \\
 ${b}_{17}=-0.015$ &${b}_{18}=0.103$& ${b}_{19}=0.130$\\
 \bottomrule
\end{tabular}
\caption{The values of  coefficients $\{b_{j}\}_{j=1,2,...19}$ in the relation function $(\ref{3.34})$.}\label{T7}%
\end{table}
In general, a connection in a $3d$ system requires  two surfaces in an implicit function description. As demonstrated here, they could be conveniently chosen if we have some knowledge about the bifurcation sequence. In contrast to previous investigations~\cite{16_1998Analytic,18_2019The}, here we take a geometric point of view and avoid employing long series to express the approximate solution. At the same time, the bifurcation parameter $r$ could be efficiently estimated through a minimization of the error function.
\subsection*{\textbf{{{Application D: Kuramoto-Sivashinsky equation}}}}\label{subsec3.4}
The Kuramoto-Sivashinsky equation(KSe) was first introduced in\cite{45_kuramoto1976persistent,46_1976Turbulent},  in which the authors studied phase turbulence in reaction-diffusion systems. Sivashinsky\cite{47_1977Nonlinear} regarded KSe as an effective description  for plane flame fronts.  It also characterises falling films on an inclined surface\cite{48_1980On}. Various properties of the KSe are sketched\cite{49_1997Infinite} including the steady states where a spiraling connection is found \cite{50_1989The,51_dong2014variational}, which depicts a front between two asymptotically flat states.  In this section, we will study this heteroclinic orbit of the steady KSe.
\par
In one spatial dimension, the KSe could be written as $u_t=uu_x-u_{xx}-u_{xxxx}$. The steady state of the KSe is described by the following equation
\begin{equation}
\begin{array}{lll}
\dot{u}&=&v\\
\dot{v}&=&w\\
\dot{w}&=&1-v-\frac{1}{2}u^2,
\end{array} \label{3.37}
\end{equation}\\
after one integration of the steady state equation $uu_x-u_{xx}-u_{xxxx}=0$ and a particular choice of the integration constant. Now, $\dot{u},...$ denotes the spatial derivative $\frac{du}{dx},...$.  Note that Eq.~(\ref{3.37}) has a reversal symmetry $x\rightarrow -x$, $u\rightarrow -u$, $v\rightarrow v$, and $w\rightarrow -w$, which could be profitably utilized in the current investigation.  In all previous calculations, we need to expand the solution at both ends of the connection. If symmetry is present, however,  only one expansion is required since the other half is recovered by symmetry. Eq.~(\ref{3.37}) has two fixed points $P_1=(\sqrt{2},0,0)$ and $P_2=(-\sqrt{2},0,0)$, being symmetry image of each other.  To locate the connection, we choose one of them, say, $P_1$ to carry out the expansion.
\begin{figure}[hbtp]
\centering
\begin{minipage}{1.5in}
\centering
\includegraphics[ width=1.5in]{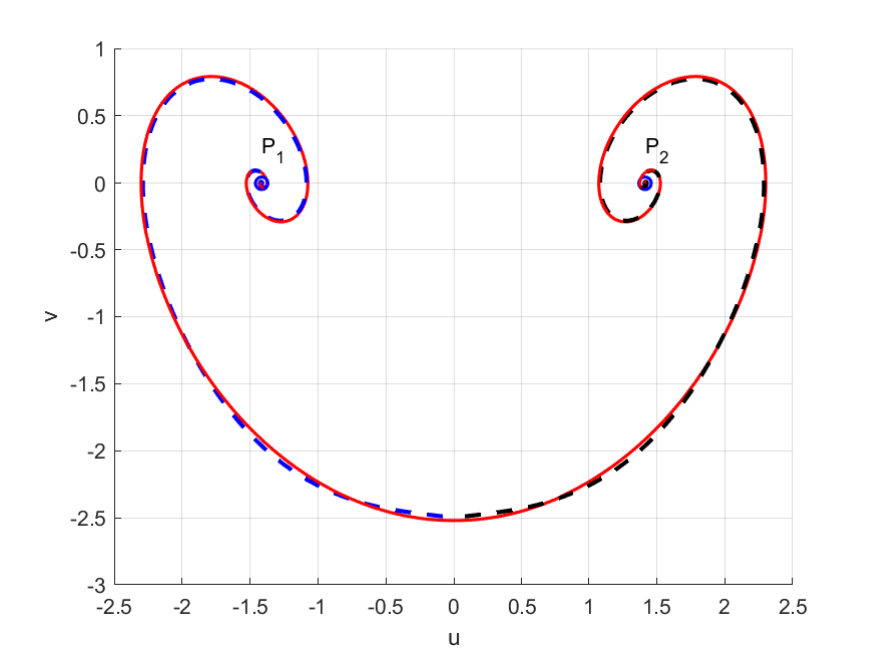}
\end{minipage}
\caption{Comparison the connecting orbit of Eq.~(\ref{3.37}) from the implicit function (blue dashed line) and its symmetrical extension (black dashed curve) with the benchmark (red solid line). }
\label{Fig6}
\end{figure}

At equilibrium $P_1$, a linearization leads to two conjugate eigenvalues $k_1=0.417+1.234i$, $k_2=0.417-1.234i$ and a negative eigenvalue $k_3=-0.834$. With the same considerations mentioned above, we only expand along the submanifolds determined by $k_1$ and $k_2$. Then, the  expansion at fixed point $P_1$ is
\begin{equation}
\begin{array}{lll}
u&=\sqrt{2}+\epsilon(h_{11}a+h_{12}b)+\epsilon^2(h_{13}a^2+h_{14}b^2+h_{15}ab)+...\\
v&=\epsilon(h_{21}a+h_{22}b)+\epsilon^2(h_{23}a^2+h_{24}b^2+h_{25}ab)+...\\
w&=\epsilon(h_{31}a+h_{32}b)+\epsilon^2(h_{33}a^2+h_{34}b^2+h_{35}ab)+...,
\end{array} \label{3.38}
\end{equation}
where $\frac{da}{dx_0}=k_1a$ and $\frac{db}{dx_0}=k_2b$.  $\{h_{jk}\}_{j,k=1,2...}$ are the expansion coefficients being listed in Table $\ref{AT5}$ of $Appendix~C$. We first determine a polynomial surface that  the orbit lies on as in the previous applications, the relation function of which may be written as $f(u,v,w)=\epsilon(a_1u+a_2v+a_3w)+\epsilon^2(a_4u^2+a_5v^2+...)+...$. However, if the connection is invariant under the symmetry operation  $u\rightarrow -u$, $v\rightarrow v$, and $w\rightarrow -w$, the relation function could be simplified to
\begin{equation}
\begin{array}{lll}
F_1(u,v,w)=&\epsilon(a_1u+a_2w)+\epsilon^2(a_3uv+a_4vw)\\
&+\epsilon^3(a_5u^3+a_6w^3+a_7uv^2+a_8u^2w\\
&+a_9v^2w+a_{10}uw^2)+\epsilon^4(a_{11}u^3v\\
&+a_{12}uv^3+a_{13}u^2vw+a_{14}v^3w\\
&+a_{15}uvw^2+a_{16}vw^3),
\end{array} \label{3.39}
\end{equation}\\
where the coefficients $\{a_{i}\}_{i=1,2,...,16}$ are determined through a  substitution of $(\ref{3.38})$ into $(\ref{3.39})$ and comparing the  terms of $a$, $b$, $a^2$,...,$a^4$, $b^4$, $a^3b$, $ab^3$, and $a^2b^2$.  With the condition that the equilibrium $P_1$ lies on the surface and the assumption $a_1=1$, we obtained $16$ linear equations, which could be easily solved for these coefficients, the results being listed in Table $\ref{T8}$.
\begin{table}[h]
\centering
\begin{tabular}{@{}llll@{}}
\toprule
$a_{1}=1$  &  $a_{2}=0.362$  & $a_{3}=0.640$ & $a_4=0.119$  \\
$a_{5}=-0.500$ &  $a_{6}=-0.110$  & $a_{7}=0.044$ &  $a_{8}=-0.771$  \\
$a_{9}=-0.079$ &  $a_{10}=-0.4617$  & $a_{11}=0.028$ &  $a_{12}=-0.019$   \\
$a_{13}=0.128$ &  $a_{14}=-0.026$  & $a_{15}=0.102$ &  $a_{16}=0.023$   \\
\bottomrule
\end{tabular}
\caption{The values of  coefficients $\{a_{i}\}_{i=1,2,...,16}$ in the relation function $(\ref{3.39})$.}\label{T8}%
\end{table}\par
 Next, we will design a second surface as done in the Lorenz case, but now at both ends there are spirals. Again, the above-mentioned symmetry is important in that only half job needs to be done. To select the correct orbit from the infinitely many spirals described by Eq.~(\ref{3.38}), information of one more point is needed. Still because of the symmetry, we take the midpoint of the connection which has the property that $u=0$ and $w=0$. By making $u=0$ and $w=0$ in approximation $(\ref{3.38})$, $a=-0.5674+0.7620i$ and $b=-0.5674-0.7620i$ is computed which is then substituted into the expression of  $v$, resulting in  $v=-2.511$. Thus, the midpoint of the heteroclinic orbit is $(u,v,w)=(0,-2.511,0)$. As before, we may write  $a=d^{\alpha}e^{i\theta}$ and $b=d^{\bar{\alpha}}e^{-i\theta}$, where $\alpha=1+2.958i$ and $\bar{\alpha}$ is the conjugate of $\alpha$. Similar to what has been done in treating spirals, $\theta=2.431$ can be estimated.  Then, the second relation function is derived from the equality  $a^{\bar{\alpha}}e^{-i\theta}=b^{\alpha}e^{i\theta}$
\begin{equation}
\begin{array}{lll}
F_2(\tilde{u},\tilde{v})&=\epsilon \tilde{u}^{\bar{\alpha}}e^{-i\bar{\alpha}\theta}(c_1+\epsilon(c_2\tilde{u}+c_{3}\frac{\tilde{v}^2}{\tilde{u}}+c_4\tilde{v}))\\
&+\epsilon \tilde{v}^{\alpha}e^{i\alpha\theta}(\bar{c}_1+\epsilon(\bar{c}_2\tilde{v}+\bar{c}_{3}\frac{\tilde{u}^2}{\tilde{v}}+\bar{c}_4\tilde{u})),
\end{array} \label{3.40}
\end{equation}
where $\{c_j,\bar{c}_{j}\}_{j=1,2,3,4}$ are the coefficients to be determined. The variables $\tilde{u}$ and $\tilde{v}$ are directly given by the following coordinate transformation
\begin{equation}
\begin{array}{lll}
\tilde{u}=g_{10}+g_{11}u+g_{12}v+g_{13}w\\
\tilde{v}=g_{20}+g_{21}u+g_{22}v+g_{23}w\\
\tilde{w}=g_{30}+g_{31}u+g_{32}v+g_{33}w,
\end{array} \label{3.41}
\end{equation}
where $\{g_{jk}\}_{j,k=1,2...}$ are known coefficients displayed in Table $\ref{AT6}$ of $Appendix~C$.  We need 8 equations to solve for the unkown coefficients in Eq.~(\ref{3.40}). Substituting the expansion $(\ref{3.38})$ and $a=d^\alpha e^{i\theta}$ and $b=d^{\bar{\alpha}}e^{-i\theta}$ into the spiral relation function $(\ref{3.40})$ together with the transformation $(\ref{3.41})$, five linear homogeneous equations can be obtained by extracting the coefficients of the terms $d^{\alpha\bar{\alpha}}$, $d^{\alpha+\alpha\bar{\alpha}}$, $d^{\bar{\alpha}+\alpha\bar{\alpha}}$, $d^{2\alpha-\bar{\alpha}+\alpha\bar{\alpha}}$, and $d^{2\bar{\alpha}-\alpha+\alpha\bar{\alpha}}$. The midpoint $(u,v,w)=(0,-2.511,0)$ should be on the surface $F_2(\tilde{u},\tilde{v})=0$,  which provides the sixth equation. As Eq.~(\ref{3.40})is homogeneous in coefficients, we may take the seventh equation as  $\mbox{Real}(c)=1$.
\par
The last equation is given by the invariance of the surface at the mid-point
\begin{equation}
\begin{array}{lll}
\frac{dF_2}{dt}=F_{2u}\dot{u}+F_{2v}\dot{v}+F_{2w}\dot{w}=0
\end{array} \label{3.42}
\end{equation}
where $F_{2u}=\frac{\partial F_2}{\partial u}$, $F_{2v}=\frac{\partial F_2}{\partial v}$, and $F_{2w}=\frac{\partial F_2}{\partial w}$. The values of $\dot{u}$, $\dot{v}$, and $\dot{w}$ are provided by the original Eq.~(\ref{3.37}) since all the coordinates are known at the mid-point $(u,v,w)=(0,-2.511,0)$.  After these eight equations are solved, the coefficients of $\{c_j,\bar{c}_{j}\}_{j=1,2,3,4}$ are derived and displayed in Table $\ref{T9}$.
\begin{table}[h]
\centering
\begin{tabular}{@{}ll@{}}
\toprule
$c_{1}=i$  & $\bar{c}_{1}=-i$       \\
 $c_{2}=0.335+0.461i$&  $\bar{c}_{2}=0.335-0.461i$   \\
 $c_3=0.082-0.052i$& $\bar{c}_3=0.082+0.052i$ \\
  $c_{4}=-0.215+0.038i$  & $\bar{c}_{4}=-0.215-0.038i$\\
  \bottomrule
\end{tabular}
\caption{The values of  coefficients $\{c_j,\bar{c}_{j}\}_{j=1,2,3,4}$ in the relation function $(\ref{3.40})$.}\label{T9}%
\end{table}
The semiorbit from the equilibria $P_1$ is displayed in Fig.$\ref{Fig6}$ in blue dashed line together with its symmetry partner (black dashed line). They match extremely well with the benchmark numerical solution plotted with  red solid line.

\subsection*{\textbf{{{Application E: A $3d$ System with exact homoclinic solution}}}}\label{subsec3.5}
If applied well, the current scheme may lead to exact solutions. Here we give such an example in $3d$ phase space. The system is adopted from the literature \cite{52_2007Nonlinear} and is given below
\begin{equation}
\begin{array}{lll}
\dot{x}&=&-\frac{1}{2}x-y+\frac{1}{2}z+2y^3\\
\dot{y}&=&-\frac{1}{2}x+\frac{1}{2}z\\
\dot{z}&=&z-xy^2+zy^2\,,
\end{array} \label{3.44}
\end{equation}
which has an obvious symmetry $x\to -x\,, y\to -y\,, z\to -z$. The origin is the symmetry center which is also an equilibrium point. The Jacobian there has three eigenvalues $k=-1$, $k_1=1$, and $k_2=\frac{1}{2}$.  There is a homoclinic orbit in system $(\ref{3.44})$ connected to the origin. As done before, an expansion along the stable manifold gives
\begin{equation}
\begin{array}{lll}
x&=a-a^3+\frac{3}{8}a^5...\\
y&=a-\frac{1}{4}a^3+\frac{1}{16}a^5...\\
z&=\frac{1}{2}a^3-\frac{1}{4}a^5+\frac{3}{32}a^7...,
\end{array} \label{3.45}
\end{equation}
and the other one is along the $2d$ unstable manifold
\begin{equation}
\begin{array}{lll}
x&=c-\frac{7}{5}c^3+\frac{8 }{3}b c^2-\frac{27}{14} b^2 c+\frac{1}{2}b^3\\
y&=b-c-\frac{1}{4}b^3+\frac{11}{14} b^2 c-\frac{2 }{3}b c^2-\frac{1}{5}c^3\\
z&=2 b-b^3+2 b^2 c-2 c^3,
\end{array} \label{3.46}
\end{equation}
where the time dependence $a$, $b$, and $c$ can be easily derived with the RG equation, $\frac{da}{dt_0}=ka$, $\frac{db}{dt_0}=k_1b$, and $\frac{dc}{dt_0}=k_2c$. Therefore, it is not difficult to obtain $c=rb^{\frac{1}{2}}$, where the parameter $r$ can be used to select the filament from the embedding unstable manifold. The first relation function is easy to choose
\begin{equation}
\begin{array}{lll}
f_1(x,y,z)&=\epsilon(a_1x+a_2y+a_3z)+\epsilon^2(a_4x^2+a_5y^2\\
&+a_6z^2+a_7xy+a_8xz+a_9yz),
\end{array} \label{3.47}
\end{equation}
where $\{a_i\}_{i=1,2,...9}$ are coefficients to be determined as presented before. This relation function is truncated at $\epsilon^3$ after a couple of trials. Substituting the above two expansions into this relation function and comparing the coefficients of the terms $a$, $a^2$, $a^3$, $a^4$, $b^{\frac{1}{2}}$,  $b$, $b^{\frac{3}{2}}$, and $b^{2}$ to zero, we get eight linear equations. However, this linear set contains the parameter  $r$ which appears in a nonlinear form.

As usual, we may take $a_1=1$ and the coefficient of $a$ gives $2+a_2=0$ while that of $b^{\frac{1}{2}}$ delivers $r-a_2r=0$. Obviously it is necessary that $r=0$, which implies that if  system $(\ref{3.44})$ has a homoclinic orbit in the neighbourhood  of the origin, it can only depart in the direction corresponding to the eigenvalue $k_1$. The coefficients are obtained as $a_2=-2$, $a_3=1$, $a_5=4a_4-2(2a_8+a_9)$, $a_6=a_8-a_4$, $a_7=-4a_4+2a_8+a_9$ with $a_4$, $a_8$, $a_9$ being free parameters. Interestingly, in this case, the relation function $(\ref{3.47})$ becomes $(x-2y+z)(1+a_9y+a_4(x-2y-z)+a_8(2y+z))$. It is convenient to try the simple form $f_1(x,y,z)=x-2y+z$. Is the plane $f_1(x,y,z)=0$ really invariant with the original dynamics? We may check its change along an orbit
\begin{equation}
\begin{array}{lll}
\frac{df_1}{dt}=f_{1x}\dot{x}+f_{1y}\dot{y}+f_{1z}\dot{z}=(x-2y+z)(\frac{1}{2}-y^2) \,,
\end{array} \label{3.48}
\end{equation}
Therefore, $f_1=0$ implies $df_1/dt=0$ and $f_1=0$ is thus an invariant plane of the system.
\par
If the other relation function is also expressed as a polynomial of $x,y,z$, then because of the first relation $f_1=0$, we may only employ two variables to represent the second surface, which is required to be invariant only at the intersection. So, we proceed to design the second relation function to be
\begin{equation}
\begin{array}{lll}
f_2(x,y)&=\epsilon(b_1x+b_2y)+\epsilon^2(b_3x^2+b_4y^2\\
&+b_5xy)+\epsilon^3(b_6x^3+b_7y^3+b_8x^2y\\
&+b_9xy^2)+\epsilon^4(b_{10}x^4+b_{11}y^4\\
&+b_{12}x^3y+b_{13}xy^3+b_{14}x^2y^2) \,,
\end{array} \label{3.49}
\end{equation}
which involves only $x,y$. The coefficients $\{b_{j}\}_{j=1,2,...}$ are determined by substituting expansions $(\ref{3.45})$ and $(\ref{3.46})$ into $(\ref{3.49})$ and comparing different orders of $a$ and $b$. Explicitly, we have $b_3=-1$,  $b_5=2$, and $b_{11}=-1$ and the remaining coefficients are all zero.  Then  the polynomial $f_2$ is $f_{2}(x,y)=2xy-x^2-y^4$ and its time rate
\begin{equation}
\begin{array}{lll}
\frac{df_2}{dt}=(y-2y^3)(x-2y+z) \,.
\end{array} \label{3.50}
\end{equation}
Therefore, it is not invariant in general but indeed invariant at the intersection with the plane $f_1=0$. Hence, the exact connection of system $(\ref{3.44})$ is described by two equations
\begin{equation}
\begin{array}{lll}
f_1(x,y,z)&=x-2y+z=0,\\
f_2(x,y)&=2xy-x^2-y^4=0\,.
\end{array} \label{3.51}
\end{equation}
Notice that the symmetry $x\rightarrow -x$, $y\rightarrow -y$, and $z\rightarrow -z$ is preserved in the system~(\ref{3.51}).
\par
\begin{figure}[hbtp]
\centering
\begin{minipage}{5cm}
\centering
\includegraphics[height=4cm, width=5cm]{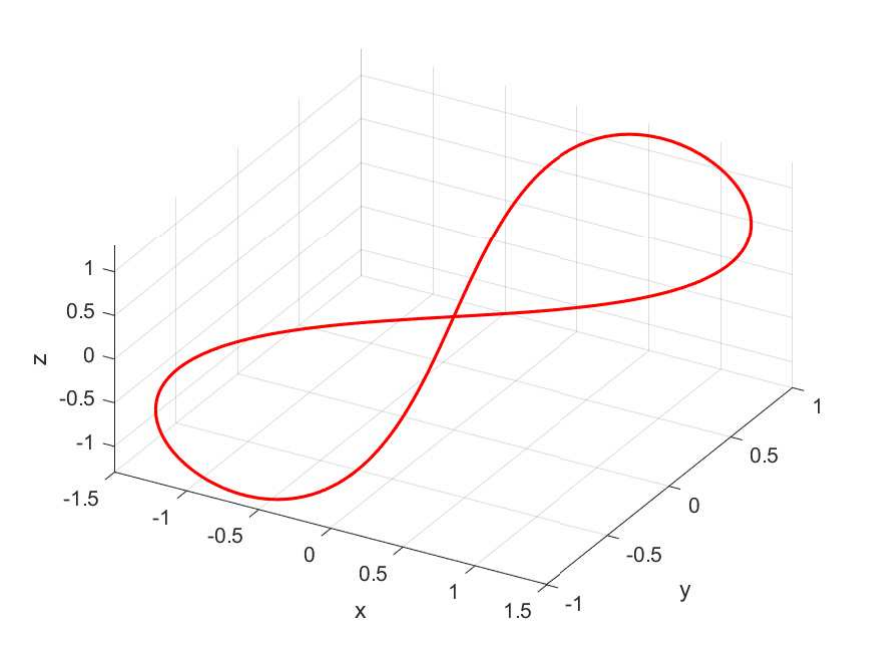}
\end{minipage}
\caption{The exact orbit depicted by the implicit functions $(\ref{3.51})$. }
\label{Fig7}
\end{figure}

This exact connection is plotted in Fig.$\ref{Fig7}$. We see a shape of "8" since there actually exist two connections because of the above-mentioned symmetry.   In the literature~\cite{52_2007Nonlinear}, this exact homoclinic solution is given in a parametric form $x=(1+\tanh{t})/\cosh{t}$, $y=1/\cosh{t}$, and $z=(1-\tanh{t})/\cosh{t}$, which only describes one orbit with $x>0,y>0,z>0$. Its symmetric partner with $x<0,y<0,z<0$ is obtained by symmetry reflection.
\section{\textbf{Conclusions}}\label{sec4}\par
Various techniques have been designed to compute connecting orbits in nonlinear systems. Yet, a uniform analytic approximation of the whole orbit is still rare.  In a previous computation~\cite{35_Yueheng2013Bridging} based on the RG, we start from one equilibrium and obtain the other one approximately. For connections with complex structures, the convergence is generally bad. In the current investigation, expansions are carried out at both ends instead. Implicit functions defined by the relation functions are used to depict the orbit and accommodate these expansions. In this way, the convergence is much improved, resulting in very nice analytic representation of the orbit.
\par
The form of the relation function combines the local expressions at both ends, which is often a polynomial but could be in a generalized type with non-integer powers derived from basic relation between variables such as $a=rb^{\frac{3}{2}}$ in $(\ref{3.6})$. Only one relation function is needed in $2d$ system (Application A and B) while two functions have to be designed to represent one orbit in a $3d$ phase space (Application C, D, and E). For the design, the experience in low dimensions could be exercised (as shown in the spiral connections) and expansions at other equilibrium points may be profitably utilized (homoclinic orbit in the Lorenz system). Symmetry is very useful in determining a connection as demonstrated in Application D.
\par
One advantage of the current method is that the coefficients of the relation function satisfy a set of linear algebraic equations which is easily solvable numerically. If there are extra parameters either from the original differential system or the initial condition, they often enter in a nonlinear way but can be estimated with an error function (Application A). In all applications, the function is finite and of low order but the depicted orbit matches the benchmark very well. In some circumstances, exact solutions could also be derived (Application E).
\par
Still,  there are many issues that need further investigation. Currently, the form of the relation function is  empirical and very likely not the best one. It would be very interesting to set up a systematic way of procuring descent forms of these functions, especially in multi-dimensions. There is no proof given for the convergence of the current scheme and we do not know if error decreases when more terms are used to represent the connection. Apparently, the technique may be used to portray invariant geometric objects of higher dimensions but more consideration should be made of boundary conditions.

\section*{Acknowledgements}
This work was supported by the National Natural Science Foundation of China under Grants No.12375030.
\section*{Conflict of Interest}
		The authors declare that they have no conflict of interest.
\section*{Data Availability Statement}
		All data, models, generated or used during the study appear in the submitted article, code generated during the study are available from the corresponding author by request.	
\newpage
\section*{Appendix}
\subsection*{\textbf{{Appendix A. The supplements  of Application B: System with homoclinic bifurcation}}}\label{secAppA}\par
In Appendix A, more details about Application B: System with homoclinic bifurcation will be displayed.
\subsubsection*{\textbf{Appendix A(1). The details in Application B: Implicit function for homoclinic orbit}}\label{secAppA1}\par
Here, we will give the detailed expressions   of $\{p_i\}$ and $\{q_{i}\}_{i=1,2,...3}$ in  expansions $(\ref{3.12})$ and $(\ref{3.13})$.
\begin{equation}
\begin{array}{lll}
q_1&=\frac{k_1-1}{4k_1^2-2\mu k_1-1},&q_2=\frac{q_1(3k_1-2)}{9k_1^2-3\mu k_1-1},\\
q_3&=\frac{(q_1^2+2q_2)(2k_1-1)}{16k_1^2-4\mu k_1-1},&p_1=\frac{k_2-1}{4k_2^2-2\mu k_2-1},\\
p_2&=\frac{p_1(3k_2-2)}{9k_2^2-3\mu k_2-1},&p_3=\frac{(p_1^2+2p_2)(2k_2-1)}{16k_2^2-4\mu k_2-1}.
\end{array} \label{app1}
\end{equation}\par
\subsubsection*{\textbf{Appendix A(2). The details in Application B: Implicit function for heteroclinic orbit with single spiral structure}}\label{secAppA2}\par
Here we will explain the procedure to get $\theta=5.320$. \par
\begin{figure}[hbtp]
\centering
\begin{minipage}{1.5in}
\centering
\includegraphics[width=1.5in]{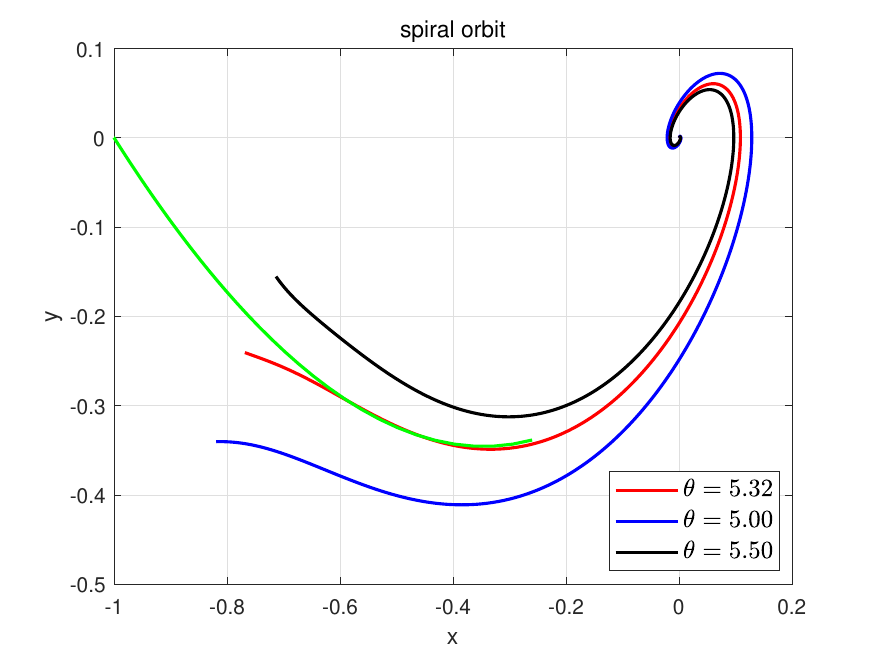}
\end{minipage}
\caption{The different trajectories when we take different values of $\theta$.}
\label{TFig1}
\end{figure}

We have obtained the  approximations near the origin and the equilibria $P_2$  presented as $(\ref{3.15})$ and $(\ref{3.16})$. Substituting the transformation $(\ref{3.150})$ into $(\ref{3.15})$ and its numerical result in $x,y$ labels  are shown in Fig.$\ref{TFig1}$ depicted by green line. Notice that the green trajectory is unique since the expression $(\ref{3.15})$ does not depend on $\theta$. The fact is that curve depicted by  expansion $(\ref{3.16})$ should connect the green curve with one suitable $\theta$ because  the two expansions represent  the same heteroclinic orbit.   Then,  we find that the numerical result of expansion near $P_2$ could be approximately connected the green line when  $\theta=5.32+2n\pi$, $n \in \mathbb{Z}$(red spiral curve in Fig.$\ref{TFig1}$) , but not connected when taking other values of $\theta$ such as $\theta=5.5$(black spiral) and $\theta=5.0$(blue spiral). Therefore, we take the singular estimation of  $\theta=5.32$ in general.  And the operation here is a more straightforward alternative application of the error function idea.
\par
\newpage
\subsection*{\textbf{{Appendix B. The  coefficients of Application C: the Lorenz System}}}\label{secAppB}\par
We  list the  coefficients mentioned in Application C. \par
\subsection*{\textbf{Appendix B$(1)$. The  coefficients in Application C:  Implicit function for heteroclinic orbit with single spiral structure for
Lorenz system }}\label{secAppB1}\par
$1$. The coefficients $\{g_{ij}\}_{i=1,2,3,j=0,1,2,3}$ in $(\ref{3.19})$ are shown in Table $\ref{AT1}$\par
\begin{table}[h]
\begin{tabular}{@{}lll@{}}
\toprule
$g_{10}=-1.364+2.169i$&$g_{11}=0.092-0.080i$&$g_{12}=-0.316-0.838i$ \\
$g_{13}=0.524+0.207i$ &$g_{20}=-1.364-2.169i$&$g_{21}=0.092+0.080i$ \\
$g_{22}=-0.316+0.838i$&$g_{23}=0.524-0.207i$&$g_{30}=-1.271$   \\
$g_{31}=-0.185$      &$g_{32}=0.632$   &     $g_{33}=-0.048$     \\
\bottomrule
\end{tabular}
\caption{The  coefficients $\{g_{ij}\}_{i=1,2,3,j=0,1,2,3}$ in $(\ref{3.19})$.}\label{AT1}%
\end{table}\par
$2$. The coefficients $\{p_{ij}\}_{i=1,2,3,j=0,1,2,3}$ in $(\ref{3.20})$ are shown in Table $\ref{AT2}$\par
\begin{table}[h]
\begin{tabular}{@{}lll@{}}
\toprule
$p_{11}=0.107+0.168i$&$p_{12}=0.042+ 0.088i$&$p_{13}=-0.137+0.211i$ \\
$p_{21}=0.042-0.088i$&$p_{22}=0.107-0.168i$&$p_{23}=-0.137-0.211i$ \\
$p_{31}=-0.084+0.013i$&$p_{32}=-0.084-0.013i$&$p_{33}=0.115$       \\
\bottomrule
\end{tabular}
\caption{The  coefficients $\{p_{ij}\}_{i=1,2,3,j=0,1,2,3}$ in $(\ref{3.20})$.}\label{AT2}%
\end{table}\par
$3$. The coefficients $\{u_j,v_j,w_{j}\}_{j=1,2,3}$  in $(\ref{3.21})$ are shown in Table $\ref{AT3}$\par
\begin{table}[h]
\begin{tabular}{@{}lll@{}}
\toprule
$u_{1}=-1.364+2.169i$&$u_{2}=-0.244-0.901i$&$u_{3}=0.048+0.019i$ \\
$v_{1}=-1.364-2.169i$&$v_{2}=-0.244+0.901i$&$v_{3}=0.048-0.019i$ \\
$w_{1}=-1.272$&$w_{2}=0.489$&$w_{3}=-0.004$       \\
\bottomrule
\end{tabular}
\caption{The  coefficients $\{u_j,v_j,w_{j}\}_{j=1,2,3}$ in $(\ref{3.21})$.}\label{AT3}%
\end{table}\par
\subsubsection*{\textbf{Appendix B$(2)$. The  coefficients in Application C: Implicit function for homoclinic orbit for the Lorenz system}}\label{secAppB2}
$1$. The coefficients $\{n_{jk}\}_{j,k=1,2,3}$  in $(\ref{3.28})$ are listed  as
\begin{widetext}
\begin{equation}
\begin{array}{lll}
n_{11}&=\frac{3\lambda_3(10+\lambda)}{(4+3\lambda_3)(\lambda_3-\lambda_1)(10+33\lambda_3+9\lambda_3^2-10r)},\\
n_{12}&=-\frac{9 \lambda_3 \left(9 \lambda_3^3+160 \lambda_3^2+760 \lambda_3+600\right)}{5 (3 \lambda_3+2) (3 \lambda_3+4)^2 (\lambda_3-\lambda_1) (9 \lambda_3^2+33 \lambda_3-10 r+10) (5 \lambda_3^2+11 \lambda_3-2 r+2)},\\
n_{13}&=\frac{81 \lambda_3 (\lambda_3+10) \left(2106 \lambda_3^6+38451 \lambda_3^5+\lambda_3^4 (236458-1620 r)+\lambda_3^3 (612500-24060 r)\right)}{10 (3 \lambda_3+2) (3 \lambda_3+4)^3 (9 \lambda_3+4) (\lambda_3-\lambda_1) \left(9 \lambda_3^2+33 \lambda_3-10 r+10\right)^2 \left(49 \lambda_3^2+77 \lambda_3-10 r+10\right) \left(5 \lambda_3^2+11 \lambda_3-2 r+2\right)},\\
&+\frac{81 \lambda_3 (\lambda_3+10) \left(-80 \lambda_3^2 (1311 r-8570)-400 \lambda_3 (331 r-793)-48000 (r-1)\right)}{10 (3 \lambda_3+2) (3 \lambda_3+4)^3 (9 \lambda_3+4) (\lambda_3-\lambda_1) \left(9 \lambda_3^2+33 \lambda_3-10 r+10\right)^2 \left(49 \lambda_3^2+77 \lambda_3-10 r+10\right) \left(5 \lambda_3^2+11 \lambda_3-2 r+2\right)},\\
n_{21}&=\frac{3 (\lambda_3+10)}{60 \lambda_3+80},\\
n_{22}&=-\frac{9 \left(\lambda_3^2+15 \lambda_3+50\right)}{20 (3 \lambda_3+2) (3 \lambda_3+4) \left(9 \lambda_3^2+33 \lambda_3-10 r+10\right)},\\
n_{23}&=\frac{27 \left(3 \lambda_3^2+40 \lambda_3+100\right) \left(237 \lambda_3^4+2819 \lambda_3^3+\lambda_3^2 (8170-210 r)-20 \lambda_3 (86 r-339)-1400 (r-1)\right)}{400 (3 \lambda_3+2) (3 \lambda_3+4)^2 (9 k+4) \left(9 \lambda_3^2+33 \lambda_3-10 r+10\right)^2 \left(5 \lambda_3^2+11 \lambda_3-2 r+2\right)},\\
n_{31}&=-\frac{3 (\lambda_3+10) (3 \lambda_3-\lambda_1)}{2 (3 \lambda_3+4) (\lambda_3-\lambda_1) \left(9 \lambda_3^2+33 \lambda_3-10 r+10\right)},\\
n_{32}&=\frac{9 \left(9 \lambda_3^3+160 \lambda_3^2+760 \lambda_3+600\right) (5 \lambda_3-\lambda_1)}{20 (3 \lambda_3+2) (3 \lambda_3+4)^2 (\lambda_3-\lambda_1) \left(9 \lambda_3^2+33 \lambda_3-10 r+10\right) \left(5 \lambda_3^2+11 \lambda_3-2 r+2\right)},\\
n_{33}&=\frac{27 (\lambda_3+10) (\lambda_1-7 \lambda_3) \left(2106 \lambda_3^6+38451 \lambda_3^5+\lambda_3^4 (236458-1620 r)+\lambda_3^3 (612500-24060 r)\right)}{20 (3 \lambda_3+2) (3 \lambda_3+4)^3 (9 \lambda_3+4) (\lambda_3-\lambda_1) \left(9 \lambda_3^2+33 \lambda_3-10 r+10\right)^2 \left(49 \lambda_3^2+77 \lambda_3-10 r+10\right) \left(5 \lambda_3^2+11 \lambda_3-2 r+2\right)},\\
&+\frac{27 (\lambda_3+10) (\lambda_1-7 \lambda_3) \left(-80 \lambda_3^2 (1311 r-8570)-400 \lambda_3 (331 r-793)-48000 (r-1)\right)}{20 (3 \lambda_3+2) (3 \lambda_3+4)^3 (9 \lambda_3+4) (\lambda_3-\lambda_1) \left(9 \lambda_3^2+33 \lambda_3-10 r+10\right)^2 \left(49 \lambda_3^2+77 \lambda_3-10 r+10\right) \left(5 \lambda_3^2+11 \lambda_3-2 r+2\right)}.
\end{array} \label{app2}
\end{equation}
\end{widetext}\par
$2$. The coefficients $\{m_{jk}\}_{j,k=1,2,...}$  in $(\ref{3.29})$ are listed  as
\begin{widetext}
\begin{equation}
\begin{array}{lll}
m_{11}&=-\frac{10 (-\lambda_1-\lambda_2+\lambda_3)}{(\lambda_3-\lambda_1) \left(\lambda_1^2+\lambda_1 (2 \lambda_2+11)+\lambda_2^2+11\lambda_2-10 r+10\right)},\\
m_{12}&=\frac{100 (-\lambda_1-2 \lambda_2+\lambda_3)}{(\lambda_3-\lambda_1) \left(\lambda_1^2+4 \lambda_1 \lambda_2+11 \lambda_1+4 \lambda_2^2+22 \lambda_2-10 r+10\right) \left(\lambda_1^2+\lambda_1 (2 \lambda_2+11)+\lambda_2^2+11 \lambda_2-10 r+10\right)},\\
m_{13}&=-\frac{3 (\lambda_1+10) (\lambda_3-3 \lambda_1)}{2 (3 \lambda_1+4) (\lambda_3-\lambda_1) \left(9 \lambda_1^2+33 \lambda_1-10 r+10\right)},\\
m_{21}&=\frac{3 (\lambda_1+10)}{60 \lambda_1+80},\\
m_{22}&=-\frac{3 (2 \lambda_1+\lambda_2+20)}{(6 \lambda_1+3 \lambda_2+8) \left(\lambda_1^2+\lambda_1 (2 \lambda_2+11)+\lambda_2^2+11 \lambda_2-10 r+10\right)},\\
m_{31}&=-\frac{10 \lambda_2}{(\lambda_3-\lambda_1) \left(\lambda_1^2+\lambda_1 (2 \lambda_2+11)+\lambda_2^2+11 \lambda_2-10 r+10\right)},\\
m_{32}&=\frac{200 \lambda_2}{(\lambda_3-\lambda_1) \left(\lambda_1^2+4 \lambda_1 \lambda_2+11 \lambda_1+4\lambda_2^2+22\lambda_2-10 r+10\right) \left(\lambda_1^2+\lambda_1 (2 \lambda_2+11)+\lambda_2^2+11 \lambda_2-10r+10\right)},\\
m_{33}&=-\frac{3 \lambda_1 (\lambda_1+10)}{(3 \lambda_1+4) (\lambda_3-\lambda_1) \left(9 \lambda_1^2+33 \lambda_1-10 r+10\right)}.
\end{array} \label{app3}
\end{equation}
\end{widetext}\par
\subsection*{\textbf{{Appendix C. The  coefficients of Application D: the Kuramoto-Sivashinsky equation}}}\label{secAppC}\par
$1$. The coefficients $\{h_{jk}\}_{j,k=1,2,...}$  in $(\ref{3.38})$ are shown in Table $\ref{AT5}$.
\begin{table}[h]
\begin{tabular}{@{}lll@{}}
\toprule
$h_{11}=1$  &              $h_{21}=0.417 + 1.234i$     &$h_{31}=-1.348+1.029i$ \\
$h_{12}=1$  &              $h_{22}=0.417 - 1.234i$     &$h_{32}=-1.348-1.029i$ \\
$h_{13}=0.029-0.018i$ &    $h_{23}=0.069+0.059i$       &$h_{33}=-0.087+0.218i$ \\
$h_{14}=0.029+0.018i$ &    $h_{24}=0.069-0.059i$       &$h_{34}=-0.087-0.218i$ \\
$h_{15}=-0.353$       &    $h_{25}=-0.295$             &$h_{35}=-0.246$ \\
\bottomrule
\end{tabular}
\caption{The  coefficients $\{h_{jk}\}_{j,k=1,2,...}$   in $(\ref{3.38})$.}\label{AT5}%
\end{table}\\
$2$. The coefficients $\{g_{jk}\}_{j,k=1,2,...}$  in $(\ref{3.40})$ are shown in Table $\ref{AT6}$.
\begin{table}[h]
\begin{tabular}{@{}lll@{}}
\toprule
$g_{10}=-0.317+0.155i$  &  $g_{11}=0.225-0.110i$  &              $g_{21}=0.135-0.268i$     \\
 $g_{31}=0.162+0.164i$&$g_{20}=-0.317-0.155i$  &  $g_{12}=0.225+0.110i$               \\
 $g_{22}=0.135+0.268i$     &$g_{32}=0.162-0.164i$&$g_{30}=-0.777$    \\
  $g_{13}=0.549$ &    $g_{23}=-0.270$       &$g_{33}=0.324$ \\
\bottomrule
\end{tabular}
\caption{The  coefficients $\{g_{jk}\}_{j,k=1,2,...}$  in $(\ref{3.40})$.}\label{AT6}%
\end{table}


\begin{thebibliography}{10}
\expandafter\ifx\csname url\endcsname\relax
  \def\url#1{\texttt{#1}}\fi
\expandafter\ifx\csname urlprefix\endcsname\relax\def\urlprefix{URL }\fi
\expandafter\ifx\csname href\endcsname\relax
  \def\href#1#2{#2} \def\path#1{#1}\fi

\bibitem{1_1983Solitons}
D.~J. Wallace, Solitons and instantons: An introduction to solitons and
  instantons in quantum field theory, Physics Bulletin. 34~(1) (1983) 29--29,
  doi:{\color{blue} \href{https://doi.org/10.1088/0031-9112/34/1/043}
  {10.1088/0031-9112/34/1/043}}.

\bibitem{2_2000Nonlinear}
E.~Infeld, G.~Rowlands, Nonlinear Waves, Solitons and Chaos, 2000,
  doi:{\color{blue} \href{https://doi.org/10.1017/CBO9781139171281}
  {10.1017/CBO9781139171281}}.

\bibitem{3_1992Fronts}
W.~V. Saarloos, P.~C. Hohenberg, Fronts, pulses, sources and sinks in
  generalized complex ginzburg-landau equations, Physica D. 56 (1992) 303--367,
  doi:{\color{blue} \href{https://doi.org/10.1016/0167-2789(92)90175-M}
  {10.1016/0167-2789(92)90175-M}}.

\bibitem{4_zwanzig2001nonequilibrium}
Z.~Robert, Nonequilibrium statistical mechanics, Oxford University Press, 2001.

\bibitem{5_2000Nonlinear}
S.~H. Strogatz, Nonlinear dynamics and chaos, J. Stat. Phys. 78~(5-6) (2000)
  1635--1636, doi:{\color{blue} \href{https://doi.org/10.1007/BF02180148}
  {10.1007/BF02180148}}.

\bibitem{8_khibnik1993periodic}
A.~I. Khibnik, D.~Roose, L.~O. Chua, On periodic orbits and homoclinic
  bifurcations in chua's circuit with a smooth nonlinearity, Int. J. Bifurcat.
  Chaos 3~(02) (1993) 363--384, doi:{\color{blue}
  \href{https://doi.org/10.1142/S021812749300026X}
  {10.1142/S021812749300026X}}.

\bibitem{9_1963Deterministic}
E.~N. Lorenz, Deterministic Nonperiodic Flow, springer, New York, 1963.

\bibitem{10_Robinson2000Nonsymmetric}
C.~Robinson, Nonsymmetric lorenz attractors from a homoclinic bifurcation,
  SIAM. J. Math. Anal. 32~(1) (2000) 119-141, doi:{\color{blue}
  \href{https://doi.org/10.1137/S0036141098343598}
  {10.1137/S0036141098343598}}.

\bibitem{11_2013Hopf}
H.~Li, M.~Wang, Hopf bifurcation analysis in a lorenz-type system, Nonlinear
  Dynam. 71~(1-2) (2013) 235--240, doi:{\color{blue}
  \href{https://doi.org/10.1007/s11071-012-0655-0}
  {10.1007/s11071-012-0655-0}}.

\bibitem{12_Leonov2016Necessary}
G.~A. Leonov, Necessary and sufficient conditions of the existence of
  homoclinic trajectories and cascade of bifurcations in lorenz-like systems:
  birth of strange attractor and 9 homoclinic bifurcations, Nonlinear Dynam.
  84~(2) (2016) 1055--1062, doi:{\color{blue}
  \href{https://doi.org/10.1007/s11071-015-2549-4}
  {10.1007/s11071-015-2549-4}}.

\bibitem{13_champneys1993hunting}
A.~R. Champneys, A.~Spence, Hunting for homoclinic orbits in reversible
  systems: a shooting technique, Adv. Comput. Math. 1 (1993) 81--108,
  doi:{\color{blue} \href{https://doi.org/10.1007/BF02070822}
  {10.1007/BF02070822}}.

\bibitem{44_hassard1994existence}
B.~Hassard, J.~Zhang, Existence of a homoclinic orbit of the lorenz system by
  precise shooting, SIAM. J. Math. Anal. 25~(1) (1994) 179-196,
  doi:{\color{blue} \href{https://doi.org/10.1137/S0036141092234827}
  {10.1137/S0036141092234827}}.

\bibitem{14_2014A}
C.~Dong, Y.~Lan, A variational approach to connecting orbits in nonlinear
  dynamical systems, Phys. Lett. A. 378-(9) (2024) 705-712,
  doi:{\color{blue}\href{https://doi.org/10.1016/j.physleta.2014.01.001}
  {10.1016/j.physleta.2014.01.001}}.

\bibitem{15_2019Computing}
M.~Farano, S.~Cherubini, J.~C. Robinet, D.~P. Palma, T.~M. Schneider, Computing
  heteroclinic orbits using adjoint-based methods, J. Fluid. Mech. 858,
  doi:{\color{blue} \href{https://doi.org/10.1017/jfm.2018.860}
  {10.1017/jfm.2018.860}}.

\bibitem{19_1981Applications}
J.~Carr, Applications of Center Manifold Theory, Springer, New York, 2012.

\bibitem{20_1984Chemical}
Y.~Kuramoto, Chemical oscillations, waves, and turbulence, Springer, Berlin,
  1984.

\bibitem{21_1999Variational}
J.~He, Variational iteration method a kind of nonlinear analytical technique:
  some examples, Int. J. Nonlin. Mech.Doi:{\color{blue}
  \href{https://doi.org/10.1016/S0020-7462(98)00048-1}
  {10.1016/S0020-7462(98)00048-1}}.

\bibitem{16_1998Analytic}
A.~F. Vakakis, M.~F.~A. Azeez, Analytic approximation of the homoclinic orbits
  of the lorenz system at $\sigma$=10, b=8/3, and $\rho$=13.926., Nonlinear
  Dynam. 15~(3) (1998) 245--257, doi:{\color{blue}
  \href{https://doi.org/10.1023/A:1008202529152} {10.1023/A:1008202529152}}.

\bibitem{18_2019The}
J.~Song, Y.~Niu, X.~Li, The existence of homoclinic orbits in the lorenz system
  via the undetermined coefficient method, Appl. Math. Comput. 355,
  doi:{\color{blue} \href{https://doi.org/10.1016/j.amc.2019.03.011}
  {10.1016/j.amc.2019.03.011}}.

\bibitem{17_2012Existence}
M.~M. El-Dessoky, M.~T. Yassen, E.~Saleh, E.~S. Aly, Existence of heteroclinic
  and homoclinic orbits in two different chaotic dynamical systems, Appl. Math.
  Comput. 218~(24) (2012) 11859--11870, doi:{\color{blue}
  \href{https://doi.org/10.1016/j.amc.2012.05.048}
  {10.1016/j.amc.2012.05.048}}.

\bibitem{35_Yueheng2013Bridging}
Y.~Lan, Bridging steady states with renormalization group analysis, Phys. Rev.
  E. 87~(1) (2013) 12914--12914, doi:{\color{blue}
  \href{https://doi.org/10.1103/PhysRevE.87.012914}
  {10.1103/PhysRevE.87.012914}}.

\bibitem{24_Wilson1971Renormalization}
K.~G. Wilson, Renormalization group and critical phenomena.
  \uppercase\expandafter{\romannumeral1}. renormalization group and the
  kadanoff scaling picture, Phys. Rev. B. 4~(9) (1971) 3174--3183,
  doi:{\color{blue} \href{https://doi.org/10.1103/PhysRevB.4.3174}
  {10.1103/PhysRevB.4.3174}}.

\bibitem{25_1971Renormalization}
K.~G. Wilson, Renormalization group and critical phenomena.
  \uppercase\expandafter{\romannumeral2}. phase space cell analysis of critical
  behavior, Phys. Rev. B. 4~(9) (1971) 3184--3205, doi:{\color{blue}
  \href{https://doi.org/10.1103/PhysRevB.4.3184} {10.1103/PhysRevB.4.3184}}.

\bibitem{26_Wilson1972Renormalization}
K.~G. Wilson, Renormalization of a scalar field theory in strong coupling,
  Phys. Rev. D. 6-(2) (1972) 419--426, doi:{\color{blue}
  \href{https://doi.org/10.1103/PhysRevD.6.419} {10.1103/PhysRevD.6.419}}.

\bibitem{27_1989Quantum}
J.~Zinnjustin, Quantum Field Theory and Critical Phenomena, Oxford University
  Press, 1989.

\bibitem{22_1994Renormalization}
L.~Chen, N.~Goldenfeld, Y.~Oono, Renormalization group theory for global
  asymptotic analysis, Phys. Rev. Lett. 74~(10), doi:{\color{blue}
  \href{https://doi.org/10.1103/PhysRevLett.74.1889.2}
  {10.1103/PhysRevLett.74.1889.2}}.

\bibitem{23_1995The}
L.~Chen, N.~Goldenfeld, Y.~Oono, The renormalization group and singular
  perturbations: Multiple scales, boundary layers and reductive perturbation
  theory, Phys. Rev. E. 54~(1) (1995) 376, doi:{\color{blue}
  \href{https://doi.org/10.1103/physreve.54.376} {10.1103/physreve.54.376}}.

\bibitem{28_bricmont1992renormalization}
J.~Bricmont, A.~Kupiainen, Renormalization group and the ginzburg-landau
  equation, Commun. Math. Phys. 150 (1992) 193--208, doi:{\color{blue}
  \href{https://doi.org/10.1007/BF02096573} {10.1007/BF02096573}}.

\bibitem{29_bricmont1994renormalization}
J.~Bricmont, A.~Kupiainen, G.~Lin, Renormalization group and asymptotics of
  solutions of nonlinear parabolic equations, Commun. Pur. Appl. Math. 47~(6)
  (1994) 893--922, doi:{\color{blue}
  \href{https://doi.org/10.1002/cpa.3160470606} {10.1002/cpa.3160470606}}.

\bibitem{34_kunihiro2022geometrical}
T.~Kunihiro, Y.~Kikuchi, K.~Tsumura, Geometrical Formulation of
  Renormalization-Group Method as an Asymptotic Analysis, Springer, Singapore,
  2022.

\bibitem{33_Chiba2008Approximation}
H.~Chiba, Approximation of center manifolds on the renormalization group
  method, J. Math. Phys. 49~(10) (2008) 1311, doi:{\color{blue}
  \href{https://doi.org/10.1063/1.2996290} {10.1063/1.2996290}}.

\bibitem{30_2000On}
M.~Ziane, On a certain renormalization group method, J. Math. Phys. 41~(5)
  (2000) 3290--3299, doi:{\color{blue} \href{https://doi.org/10.1063/1.533307}
  {10.1063/1.533307}}.

\bibitem{31_deville2008analysis}
R.~E.~L. DeVille, A.~Harkin, M.~Holzer, K.~Josi{\'c}, T.~J. Kaper, Analysis of
  a renormalization group method and normal form theory for perturbed ordinary
  differential equations, Phys. D. 237~(8) (2008) 1029--1052, doi:{\color{blue}
  \href{https://doi.org/10.1016/j.physd.2007.12.009}
  {10.1016/j.physd.2007.12.009}}.

\bibitem{32_chiba2008c}
H.~Chiba, C$^{1}$ approximation of vector fields based on the renormalization
  group method, SIAM. J. Appl. Dyn. Syst. 7~(3) (2008) 895--932,
  doi:{\color{blue} \href{https://doi.org/10.1137/070694892}
  {10.1137/070694892}}.

\bibitem{36_kupiainen2016renormalization}
A.~Kupiainen, Renormalization group and stochastic pdes, Ann. Henri.
  Poincar{\'e} 17~(3) (2016) 497--535, doi:{\color{blue}
  \href{https://doi.org/10.1007/s00023-015-0408-y}
  {10.1007/s00023-015-0408-y}}.

\bibitem{37_2021Renormalization}
S.~Qu, W.~Li, S.~Shi, Renormalization group approach to sdes with nonlinear
  diffusion terms, Mediterr. J. Math. 18~(5), doi:{\color{blue}
  \href{https://doi.org/10.1007/s00009-021-01821-6}
  {10.1007/s00009-021-01821-6}}.

\bibitem{38_guo2024renormalization}
L.~Guo, Renormalization group method for a stochastic differential equation
  with multiplicative fractional white noise, Mathematics-Basel. 12~(3) (2024)
  379, doi:{\color{blue} \href{https://doi.org/10.3390/math12030379}
  {10.3390/math12030379}}.

\bibitem{39_goto2007renormalization}
S.~Goto, Renormalization reductions for systems with delay, Prog. Theor. Phys.
  118~(2) (2007) 211--227, doi:{\color{blue}
  \href{https://doi.org/10.1143/PTP.118.211} {10.1143/PTP.118.211}}.

\bibitem{40_2023Renormalization}
Z.~Xu, L.~Xu, W.~Li, S.~Shi, Renormalization group method for singular
  perturbation initial value problems with delays, Mediterr.J. Math. 20~(2),
  doi:{\color{blue} \href{https://doi.org/10.1007/s00009-023-02281-w}
  {10.1007/s00009-023-02281-w}}.

\bibitem{press2007numerical}
W.~H. Press, Numerical recipes 3rd edition: The art of scientific computing,
  Cambridge university press, 2007.

\bibitem{41_1982The}
C.~Sparrow, The Lorenz Equations: Bifurcation, Chaos, Strange Attractors,
  Vol.~41, Springer, New York, 1982.

\bibitem{42_1993New}
J.~Li, J.~Zhang, New treatment on bifurcations of periodic solutions and
  homoclinic orbits at high $r$ in the lorenz equations, SIAM. J.
  Math.Doi:{\color{blue} \href{https://doi.org/10.1137/0153053}
  {10.1137/0153053}}.

\bibitem{43_Konstantin1995Chaos}
M.~Konstantin, M.~Marian, Chaos in the lorenz equations: a computer-assisted
  proof, Bull. Amer. Math. Soc.Doi:{\color{blue}
  \href{https://doi.org/10.1090/S0273-0979-1995-00558-6}
  {10.1090/S0273-0979-1995-00558-6}}.

\bibitem{45_kuramoto1976persistent}
Y.~Kuramoto, T.~Tsuzuki, Persistent propagation of concentration waves in
  dissipative media far from thermal equilibrium, Prog. Thero. Phys. 55~(2)
  (1976) 356--369, doi:{\color{blue} \href{https://doi.org/10.1143/PTP.55.356}
  {10.1143/PTP.55.356}}.

\bibitem{46_1976Turbulent}
Y.~Kuramoto, T.~Yamada, Turbulent state in chemical reactions, Prog. Theor.
  Phys.~(2) (1976) 679--681, doi:{\color{blue}
  \href{https://doi.org/10.1143/PTP.56.679} {10.1143/PTP.56.679}}.

\bibitem{47_1977Nonlinear}
G.~I. Sivashinsky, Nonlinear analysis of hydrodynamic instability in laminar
  flames \uppercase\expandafter{\romannumeral1}. derivation of basic equations,
  Dynamics of Curved Fronts (1988) 459--488Doi:{\color{blue}
  \href{https://doi.org/10.1016/B978-0-08-092523-3.50048-4}
  {10.1016/B978-0-08-092523-3.50048-4}}.

\bibitem{48_1980On}
G.~I. Sivashinsky, D.~M. Michelson, On irregular wavy flow of a liquid film
  down a vertical plane, Prog. Theor. Phys.Doi:{\color{blue}
  \href{https://doi.org/10.1143/PTP.63.2112} {10.1143/PTP.63.2112}}.

\bibitem{49_1997Infinite}
R.~Temam, Infinite-Dimensional Dynamical System in Mechanics and Physics,
  Springer, New York, 1997.

\bibitem{50_1989The}
W.~C. Troy, The existence of steady solutions of the kuramoto-sivashinsky
  equation, J. Differ. Equations. 82~(2) (1989) 269--313, doi:{\color{blue}
  \href{https://doi.org/10.1016/0022-0396(89)90134-4}
  {10.1016/0022-0396(89)90134-4}}.

\bibitem{51_dong2014variational}
C.~Dong, Y.~Lan, A variational approach to connecting orbits in nonlinear
  dynamical systems, Phys. Lett. A. 378~(9) (2014) 705--712, doi:{\color{blue}
  \href{https://doi.org/10.1016/j.physleta.2014.01.001}
  {10.1016/j.physleta.2014.01.001}}.

\bibitem{52_2007Nonlinear}
D.~W. Jordan, Nonlinear ordinary differential equations: an introduction for
  scientists and engineers, Oxford University Press, 2007.

\end{thebibliography}

\end{document}